\documentclass{aa}
\usepackage{graphicx}
\usepackage{txfonts}
\usepackage[english]{babel}  
\usepackage{epsfig}
\usepackage{setspace}
\titlerunning{The Earth as a transiting planet}
\authorrunning{Vidal--Madjar et al.}

\begin{document}

\title{The Earth as an extrasolar transiting planet}
\subtitle{Earth's atmospheric composition and thickness \\revealed by Lunar
eclipse observations\thanks{Detailed observations as
shown in Figs. 9, 10, 11 and 12 are only available in electronic
form at the CDS via anonymous ftp to cdsarc.u-strasbg.fr
(130.79.128.5) or via \texttt{http://cdsweb.u-strasbg.fr/cgi-bin/qcat?J/A+A/}}}

   \author{A.~Vidal--Madjar\inst{1} \and L.~Arnold\inst{2} \and D.~Ehrenreich\inst{3}
   \and R.~Ferlet\inst{1} \and A.~Lecavelier~des~Etangs\inst{1}
   \and F.~Bouchy\inst{1,2} \and D.~Segransan\inst{4} \and  I.~Boisse\inst{1}
   \and G.~H\'ebrard\inst{1} \and  C.~Moutou\inst{5}
   \and J.-M.~D\'esert\inst{1,6} \and D.~K.~Sing\inst{1,7} \and R.~Cabanac\inst{8}
   \and C.~Nitschelm\inst{9} \and X.~Bonfils\inst{3} \and X.~Delfosse\inst{3}
   \and M.~Desort\inst{3} \and R.~F.~Diaz\inst{1} \and A.~Eggenberger\inst{3}
   \and T.~Forveille\inst{3} \and A.-M.~Lagrange\inst{3} \and C.~Lovis\inst{4}
   \and F.~Pepe\inst{4} \and C.~Perrier\inst{3} \and F.~Pont\inst{7}
   \and N.~C.~Santos\inst{4,10} \and S.~Udry\inst{4}
 }

   \institute{
      Institut d'Astrophysique de Paris, UMR7095 CNRS, Universit\'e Pierre \& Marie
   Curie, 98bis, boulevard Arago, 75014 Paris, France, \email{alfred@iap.fr}
         \and
   Observatoire de Haute-Provence, CNRS/OAMP, 04870
   Saint-Michel-l'Observatoire, France
          \and
   Laboratoire d'Astrophysique de Grenoble, Universit\'e Joseph
   Fourier, CNRS (UMR 5571), BP 53, 38041 Grenoble cedex 9, France
         \and
     Observatoire de Gen\`eve, Universit\'e de Gen\`eve, 51 Chemin
     des Maillettes, 1290 Sauverny, Switzerland
          \and
    Laboratoire d'Astrophysique de Marseille, Universit\'e de Provence, CNRS (UMR6110), BP 8,
    Technop\^ole Marseille \'Etoile, 13376 Marseille Cedex 12, France
             \and
    Harvard-Smithsonian Center for Astrophysics, 60 Garden Street, Cambridge, Massachusetts 02138
    USA
             \and
    School of Physics, University of Exeter, Exeter, EX4 4QL, UK
            \and
    Observatoire Midi-Pyr\'en\'ees, TBL, 57 Ave d'Azereix, 65000 Tarbes, France
            \and
            Instituto de Astronom{\'i}a, Universidad Cat{\'o}lica del Norte,
            Avenida Angamos 0610, Antofagasta, Chile
            \and
    Centro de Astrofisica, Universidade do Porto, Rua das
    Estrelas, 4150-762 Porto, Portugal
   }

   \date{}


  \abstract
   {An important goal within the quest for detecting an Earth-like extrasolar planet,
   will be to identify atmospheric gaseous bio-signatures.}
   {Observations of the light transmitted through the
   Earth's atmosphere, as for an extrasolar planet,
   will be the first important step for future comparisons. We have
   completed observations of the Earth during a lunar eclipse, a unique
   situation similar to that of a transiting planet. We aim at showing
   what species could be detected in its atmosphere at
   optical wavelengths,
   where a lot of photons are available in the masked stellar light.}
   {We present observations of the 2008 August 16 Moon eclipse
   performed with the SOPHIE spectrograph at the Observatoire de Haute-Provence (France).
   Locating the spectrograph's fibers in the penumbra of the eclipse, the Moon irradiance
   is then a mix of direct, unabsorbed Sun light and solar light that has passed through
   the Earth's atmosphere. This mixture essentially reproduces what is recorded during the transit of an
   extrasolar planet.}
   {We report here the clear detection of several Earth atmospheric compounds
   in the transmission spectra,
   such as ozone, molecular
   oxygen, and neutral sodium as well as molecular nitrogen and oxygen through the Rayleigh
   signature. Moreover, we present a method
   that allows us to derive the thickness of the atmosphere versus the wavelength
   for penumbra eclipse observations.
   We quantitatively evaluate the altitude at which the atmosphere becomes transparent for
   important species like molecular oxygen and ozone, two species thought to be
   tightly linked to the presence of life.}
   {The molecular detections presented here are an encouraging first attempt, necessary
   to better prepare for the future of extremely-large telescopes and transiting Earth-like
   planets. Instruments like SOPHIE will be mandatory when characterizing the atmospheres
   of transiting Earth-like planets from the ground and searching for bio-marker signatures.}


  \maketitle

%

\section{Introduction}
     Soon the {\it CoRoT} (a space program operated by the French Space Agency, CNES) and
{\it Kepler} (a NASA spacecraft searching for habitable planets)
missions may discover a transiting Earth-like planet.
     It will then be of prime importance to study such an exoplanet's atmospheric
     composition, in order to define its evolutionary status
     as compared to the known Earth one. Through these observations the impact
     of galactic environmental perturbations as, {\it e.g.} the passing through a dense
     interstellar cloud (\cite{vidal-madjar1978}), may be directly evaluated.
     However, the ultimate goal of these studies will be to know if life could have emerged
     elsewhere as it did once on Earth.

     It has been demonstrated, as for the exoplanet HD209458b, that it is possible
     to detect the presence of some species in the planetary atmosphere, like sodium, hydrogen,
     oxygen, and carbon (\cite{charbonneau2002}, Vidal-Madjar et~al. 2003, 2004). More
     recently, additional detections have been reported, including HI from recombination via the
     Balmer jump (\cite{ballester2007}), H$_2$ the main gaseous content via Rayleigh
     scattering (\cite{sing2008a}, 2008b, Lecavelier des Etangs et~al. 2008b), and TiO/VO
     (D\'esert et~al. 2008).
     The possible signature from H$_2$O in HD209458b (\cite{barman2007}) is now strongly
     questioned by new {\it HST} (the Hubble Space Telescope) and {\it Spitzer} (a Space Telescope,
     studying the universe in infrared) observations, revealing on the contrary
     the presence of high-altitude haze as well as CO in the atmosphere of
     HD189733b
     (\cite{ehrenreich2007}, Lecavelier des Etangs et~al. 2008a, Sing et~al. 2009,
     D\'esert et~al. 2009).
     New transiting planets are discovered from space by {\it CoRot}, launched at the end
     of 2006, and by {\it Kepler} launched in March 2009. More specifically, space observations are
     reaching high enough accuracy to enable the detection of Earth-size transiting planets
     (\cite{borde2003}, Rouan et~al. 2009, L\'eger et~al. 2009, Queloz et~al. 2009),
     some of which are possibly ``ocean-planets''
     (\cite{leger2004})
     or even similar to the telluric planets of our Solar System, at distances compatible with
     a so-called habitable zone. Obviously, the characterization of the corresponding atmospheres would be
     an exciting achievement.

Knowing that the probability to have a transiting Earth at about
one AU from its star is of the order of 0.5\%, and that the number
of available planets rises as the cube of their distance from the
Sun, Ehrenreich et~al. (2006) have shown that the number of
available transiting planets within 60~pc is similar to the number
of possible targets available if one looks at the planetary
emissions within a sphere of about 10 pc around the Sun (the
current expectations of projects like {\it TPF-I} (the NASA,
Terrestrial Planet Finder via Interferometry) or {\it Darwin} (the
ESA equivalent program). However, the present difficulty of the
transiting approach is the lack of photons ({\it i.e.} high S/N)
necessary to analyze spectra of Earth-like planets when using 2-m
class telescopes. Lecavelier des Etangs \& Ehrenreich (2005) have
shown that many hot Jupiters, hot Neptunes, ocean planets, and
other low-density planets will be observable with larger ground
based telescopes, like the future European Extremely-Large
Telescope (E-ELT).Transmission spectroscopy has already proved its
potential to probe atmospheres of extrasolar planets. It could
very well be the first method which will give access to the
atmosphere of smaller and cooler transiting planets, hopefully
Earth-like.

Observing in the visible and near UV offers a great advantage,
because of the strong opacity of some specific narrow line species
in that spectral range, and because of the often more intense
stellar flux at these wavelengths. During a transit event, only
the upper atmosphere of the planet may be seen, thus reducing the
many perturbing effects potentially present when observing at
lower resolution, {\it i.e.} at lower altitudes, like the presence
of clouds, continents, oceans etc. as shown in
\cite{arnold2002,woolf2002,arnold2008} and references therein.
Indeed, in the core of the spectral lines we will sample only the
upper atmosphere, where one can expect that the mixing of
different species has already taken place, thus revealing an
``average'' planetary atmosphere representative of overall
signatures.

\begin{table*}
\caption{Fibers positions on the Moon during and after the
eclipse. Times are for mid-exposure.}
\begin{tabular}{cccccc}
 \hline
 Date;  &  Fiber A location &  Fiber B location  &  Exposure  & BERV & Air mass \\
 Time (UT) &  longitude; latitude &  longitude; latitude  & time (s) & (km/s) & $sec (z)$ \\
 \hline
2008--08--16; 20:44:21    &  E $52.0^{\circ}$; N $9.8^{\circ}$   &  E $68.5^{\circ}$; N $12.1^{\circ}$  & 198.92 & --0.064 &  2.94  \\
2008--08--17; 00:11:56  &  E $39^{\circ}$; S $51^{\circ}$      &  E $50^{\circ}$; S $48^{\circ}$  & 2.00  & --0.408   &  1.86  \\
2008--08--17; 02:55:45   &  E $47^{\circ}$; N $20^{\circ}$        &  E $64^{\circ}$; N $22^{\circ}$  & 3.00 & --0.667   &  2.99 \\
 \hline
\end{tabular}
\label{table_position}
\end{table*}

Following the pioneering work of Sagan et~al. (1993) searching for
``life on Earth'' from remote sensing, our purpose is to observe
the Earth as a transiting planet in order to better test the
present models of Earth-like planets and thus get ready to
properly analyze future observations of real transiting Earth-like
extrasolar planets. Using the Moon as a reflective surface during
total or partial eclipses provides the needed situation which
furthermore can be exploited from the ground. We have conducted a
first attempt during the 2008 August 16 lunar eclipse with the
SOPHIE high-resolution spectrograph
(\cite{perruchot2008,bouchy2009}) at the 1.93 cm telescope of the
OHP (Observatoire de Haute Provence) observatory covering the
3\,900 to 7\,000~\AA\ spectral range at a resolving power of
R$\sim$75\,000. The choice of a high-resolution spectrograph
allows us to better identify absorbing species via their detailed
spectral signatures.

Other observations of this eclipse have been conducted
(\cite{palle2009}) over a wider wavelength range (0.36~-~2.4~$\mu
m$), but at a lower spectral resolution (R$<$1\,000), resulting in
a mainly qualitative identification of the Earth's atmospheric
compounds. Here, we will focus on quantitative results. In
Sect.~\ref{Obs}, the SOPHIE observations are described, then a
quantitative data analysis is given in Sect.~\ref{data_analysis},
followed by a direct comparison with current model predictions in
Sect.~\ref{Q_analysis}. We show that indeed, in the visible range
some of the main Earth's atmospheric compounds are identified and
evaluated.

\section{Observations}
\label{Obs}

The observations were conducted with the cross-dispersed,
environmentally stabilized echelle spectrograph SOPHIE, dedicated
to high-precision radial velocity measurements
(\cite{perruchot2008}, \cite{bouchy2009}). They were secured in
\textit{high-resolution} mode, i.e. the spectrograph was fed by a
40~$\mu$m slit and an optical scrambler located at the output of
the optical fiber, allowing a resolving power
$\lambda/\Delta\lambda=75\,000$.

Two optical fibers, 3~arcsec wide, separated by 1.86~arcmin were
used. Both were placed on the Moon and aligned along the east-west
direction.

The spectra were extracted from the detector images with the
SOPHIE pipeline, which includes localization of the orders on the
2D-images, optimal order extraction, cosmic-ray rejection,
wavelength calibration, corrections of flat-field and charge
transfer inefficiency at low signal-to-noise ratio
(\cite{bouchy2009}).

\subsection{Fibers position on the Moon}

\begin{figure}
   \centering
\resizebox{\columnwidth}{!}{\includegraphics[width=6cm]{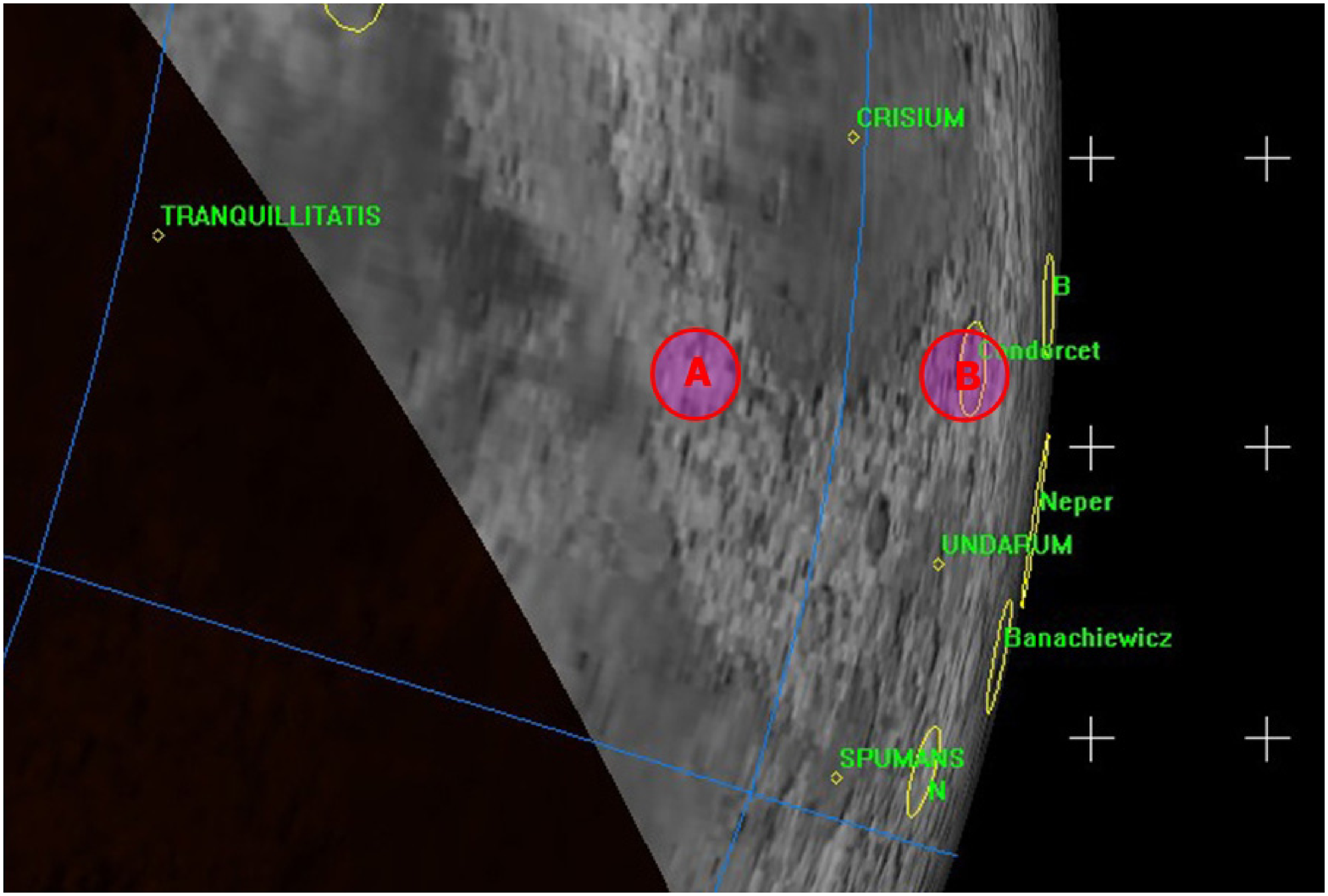}}
\resizebox{\columnwidth}{!}{\includegraphics[width=6cm]{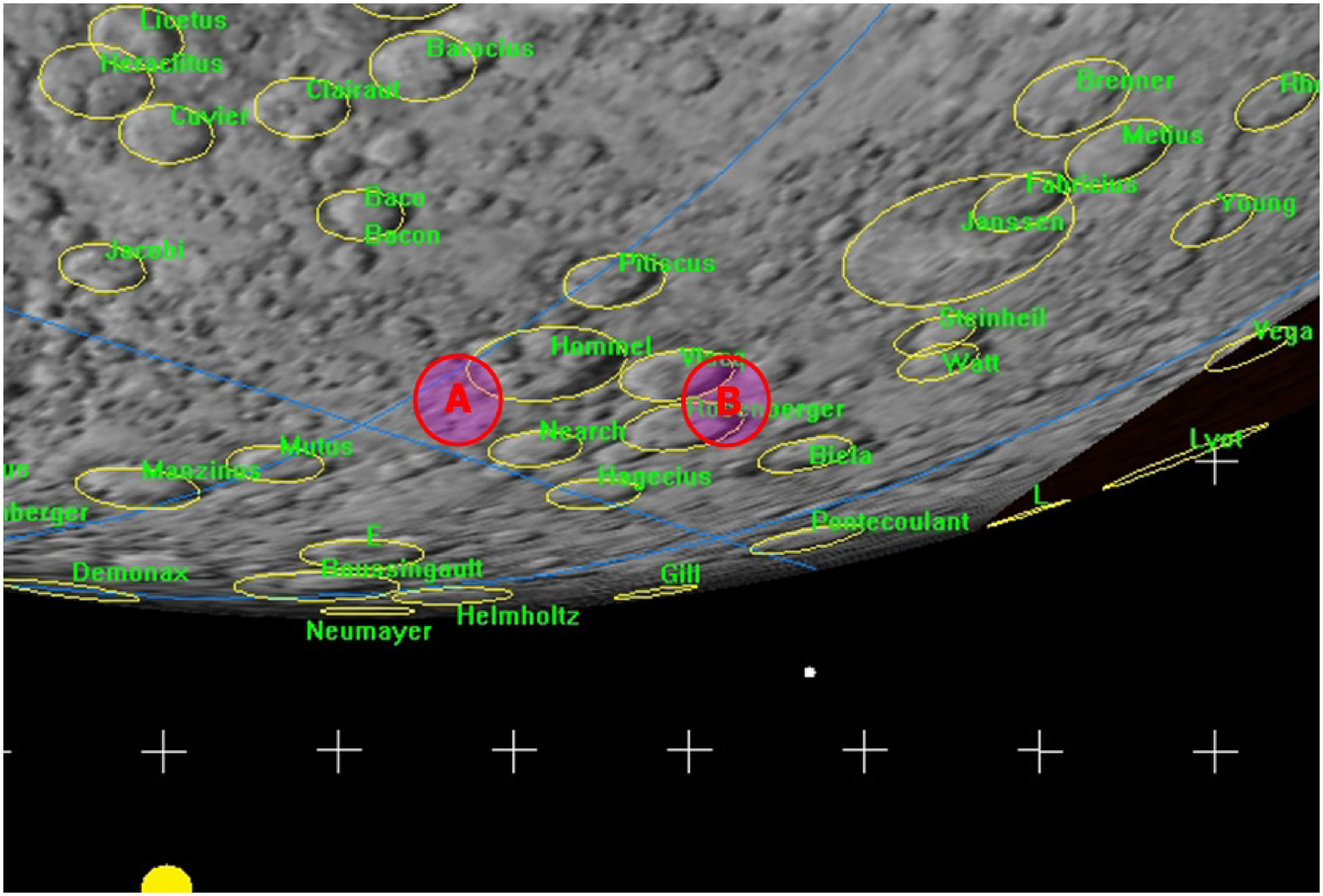}}
   \caption{Fibers A and B on the Moon in penumbra.
   The umbra is the dark area at the left (upper figure at 20h44~UT)
   and at right (lower figure at 22h44~UT).
   The distance between the fibers is 1.86 arcmin, east-west oriented.}
     \label{fibers_position}
\end{figure}

Table~\ref{table_position} lists the three main spectra used in
our analysis along with their positions over the Lunar disk as
shown in Fig.~\ref{fibers_position}. Only one observation during
the eclipse was useful, taken at 20h44~UT in the penumbra, close
to the Earth shadow. Figure~\ref{fibers_position} shows where the
fibers were positioned, {\it i.e.} over high-albedo Imbrium-type
or pre-Imbrium-type terrains (for more information see the
\cite{position} at the end of the reference list), which maximized
the flux during the eclipse. The two other observations listed in
Table~\ref{table_position} are two among a total of 13 calibration
spectra recorded after the eclipse (full Moon outside the
penumbra) at similar air masses ($sec (z)$ in
Table~\ref{table_position} where $z$ is the zenith angle) and over
the two different regions pointed to during the eclipse.
Unfortunately, after the eclipse the guiding camera was saturated
and an accurate position of the fibers could not be checked and
adjusted visually as during the eclipse. Nevertheless the
telescope coordinates indicate that the fibers were positioned
over the same kind of lunar terrains.

\subsection{Observations log}

We collected only sparse observations, because clouds appeared
just after the first exposure taken at 20h44 UT. These clouds
disappeared for a few minutes at 22h44 UT, just before the umbra
left the Moon, though these observations are more noisy and are
still possibly affected by intervening clouds.

The observations completed at 20h44 UT were done with fibers A and
B at about 2 and 4~arcmin from the penumbra/umbra limit
respectively.

Spectra were also taken later in the night after the eclipse, when
the clouds moved away. In particular at 02h56 UT spectra were
taken over the same location on the Moon and at the same air mass
as those at 20h44 UT. The spectrum taken at 00h12 UT, although
pointed over another region of the Moon, is also listed because it
was taken with a shorter time gap from the 20h44 UT eclipse
observation, leading to a smaller differential BERV correction
(``Barycentric Earth Radial Velocity'' correction of the Earth's
motion along its orbit to evaluate the spectral observations in
heliocentric coordinates as automatically done through the SOPHIE
pipeline). The corresponding BERV shifts are also listed in
Table~\ref{table_position}.

The calibration spectra (taken later in the night after the
eclipse) are used to correct the eclipse spectra from the direct
atmospheric absorption along the line of sight between the Moon
and the OHP Observatory and to extract the altitude at which the
Earth atmosphere becomes transparent enough for a grazing line of
sight.

\section{Data analysis}
\label{data_analysis}

Because we are searching for atmospheric signatures in the
gathered spectra, we have first to recalculate the wavelengths in
the geocentric reference frame. We know that there all atmospheric
signatures will be present at identical wavelengths, while the
solar line signatures will be slightly displaced by an amount
related to the spectral shift induced by the BERV. Note however
that at the SOPHIE resolution of about $\sim$4km/s, these shifts
of less than 0.5km/s are likely insignificant. The possible impact
of the BERV effect on our data is assessed in Sect.
\ref{discussion}.

Our observations present two types of Earth atmospheric absorption
signatures~: i) the ``vertical'' ones produced along the path
length between the Moon and the telescope and ii) the
``horizontal'' ones due to the solar light grazing the Earth
atmosphere before reaching the Moon. We are only interested in the
latter and will therefore have to correct the data from the
``vertical'' signatures.

Finally, knowing that the SOPHIE pipeline is optimized for high
radial velocity precisions and not for flux evaluations, we can
expect some difficulties related to flux estimates, {\it i.e.}
echelle order corrections as well as precise zero level
evaluations.

\subsection{The ``vertical'' Earth atmospheric transmission function}

The ``vertical'' atmospheric transmission is a function that
depends upon the air mass $AM$, noted $T_{AM} (\lambda)$. Along
the line of sight, an initial spectrum $I (\lambda)$ reaching the
top of the atmosphere is transformed into an observed spectrum
$O_{AM}(\lambda)$~:
\begin{equation}
  O_{AM} (\lambda) =  I (\lambda) \times T_{AM} (\lambda). \label{eqt1}
\end{equation}
For a plane parallel atmosphere (an approximation valid to air
masses smaller than $\sim$5) the air mass $AM$ is equal to
$a$=$sec(z)$ and the atmospheric transmission function is then
simply $T^a (\lambda)$, the transmission function $T$ at the power
$a$ (\cite{bird1982}), where $T$ is the transmission in the
vertical direction (air mass of 1).

Indeed for an air mass $a$ equal to the sum of two air masses
$a_1$ and $a_2$, one can evaluate the observed spectrum in the
following manner~:
\begin{equation}
  O_{a_1 + a_2}=I\times T^{a_1 + a_2}=(I\times T^{a_1}) \times T^{a_2}, \label{eqt2}
\end{equation}
showing that the transmission function $T (\lambda)$ follows the
conditions~:
\begin{equation}
  T^{a_1 + a_2} (\lambda) =  T^{a_1} (\lambda) \times T^{a_2} (\lambda) \label{eqt21}
\end{equation}
and
\begin{equation}
  T^{0} (\lambda) =  1 .\label{eqt22}
\end{equation}

If the transmission function $T (\lambda)$ is known at a given air
mass, it can be evaluated at any other air mass. Noting that

\begin{equation}
   \frac{O_{a_1} (\lambda)} {O_{a_2} (\lambda)} =  \frac{T^{a_1} (\lambda)} {T^{a_2} (\lambda)} = T^{a_1 - a_2} (\lambda),\label{eqt3}
\end{equation}
we were able to evaluate directly the transmission function at
different air masses through the complete series of spectra
gathered during the full Moon observations. We normalized all
evaluations extracted from observational ratios to an air mass
equal to 1 by calculating

\begin{equation}
   \left(\frac{O_{a_1} (\lambda)} {O_{a_2} (\lambda)}\right)^{1/(a_1 - a_2)} = (T^{a_1 - a_2} (\lambda))^{1/(a_1 - a_2)} = T (\lambda).\label{eqt31}
\end{equation}

\begin{figure}[h]
   \centering
\resizebox{\columnwidth}{!}{\includegraphics[width=6cm]{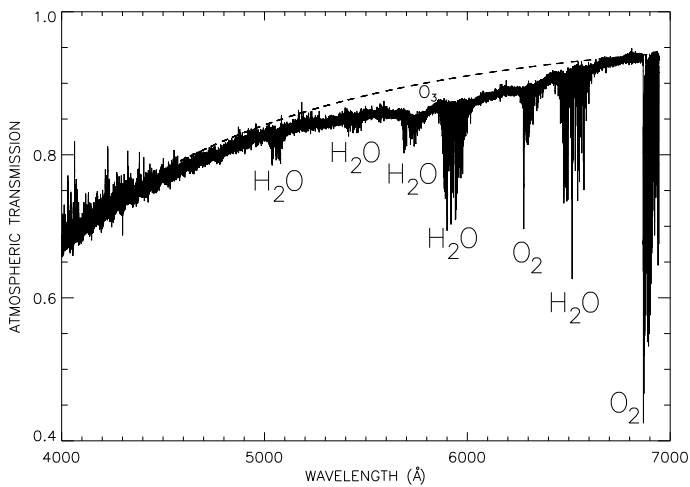}}
   \caption{Average atmospheric transmission function $T (\lambda)$ as evaluated
   during full Moon just after the eclipse supposed to be similar to the one during
   the eclipse observations. The dashed line shows a transmission model
   (\cite{hl1975}),
   which includes both Rayleigh diffusion and aerosols (optical depth of 0.035 at 5\,320~\AA\
   typical for clear OHP nights).
   The experimental evaluation was normalized in order to match
   the model at our reference wavelength $\lambda_0$~=~4\,530~\AA\
   (see Sect.~\ref{discussion}).
   O$_2$ as well as H$_2$O
   absorption bands are clearly seen. Note that the match with the model is good except
   for the central region from 5\,000 to 6\,700~\AA ,
   where a clear additional absorption is detected.
   This is the Chappuis band from ozone (noted O$_3$), not included in the model calculation.}
   \label{transmission}
\end{figure}

From the 13 full Moon observations $O_{a} (\lambda)$, we extracted
24 couples well separated in air mass each containing independent
evaluations of the transmission function $T (\lambda)$. In
Fig.~\ref{transmission}, the average of all evaluations are shown,
providing the reference transmission function $T (\lambda)$ that
we will now use in the following data analysis to re-evaluate all
observed eclipse spectra as if they were observed from outside the
atmosphere.

\subsection{Other corrections of the SOPHIE spectra}

\begin{figure}[h]
   \centering
\resizebox{\columnwidth}{!}{\includegraphics[width=6cm]{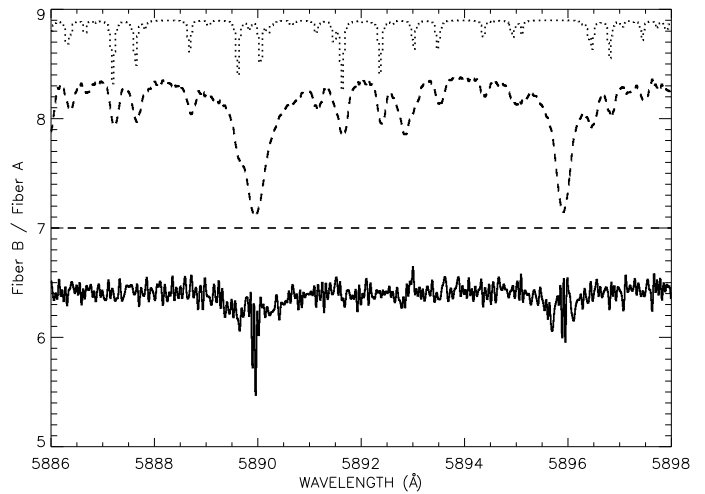}}
   \caption{
   At 20h44 UT, the (Fiber B / Fiber A) ratio (solid line)
   is shown in the NaI doublet spectral
   region. The Fiber A spectrum (dashed line) with its corresponding zero
   level
   (horizontal dashed line) as extracted from the
   SOPHIE pipeline as well as an H$_2$O model (dotted line) are overplotted
   (in arbitrary scales simply to fit within the figure frame) to
   show the positions of the solar as well as
   H$_2$O  spectral lines in the spectrum.
   Most of the observed lines are due to atmospheric
   water vapor. They disappear in the fiber's
   spectral ratio, while the broad spectral wings of the two NaI solar lines are
   still clearly visible.  }
   \label{FBsFA}
\end{figure}

 One can note in Fig.~\ref{transmission} the SOPHIE
echelle order signatures as regular wiggles over $\sim$$80$\AA\
from 4000 to 4500\AA , where they are particularly visible. Those
instrumental signatures should not be seen in an atmospheric
transmission function. These will certainly be a perturbation in
our data analysis, but for the time being we will try to analyze
the data without a modification to the SOPHIE pipeline. We will
however evaluate the impact of this approximation a posteriori by
looking how these order signatures are affecting the results (see
Sect.~\ref{discussion}).

In Fig.~\ref{FBsFA}, both the direct Fiber A spectrum observed
during the eclipse as well as the Fiber~B/Fiber~A flux ratio are
shown close to the NaI doublet solar lines. Note that the numerous
spectral lines, mostly due to H$_2$O in the Earth atmosphere,
completely disappear from the Fiber~B/Fiber~A ratio. This is not
surprising because both Fibers~A and B are observed through the
same atmospheric (and thus air mass) layer.

Although most of the weaker solar lines are also erased in this
ratio (as it can be directly seen in Fig.~\ref{transmission},
which also shows spectral ratios, see Eq.~\ref{eqt31}), for strong
and deep solar lines like {\it e.g.} the NaI lines, as shown in
Fig.~\ref{FBsFA}, the signature of their broad spectral wings is
still visible in the Fiber~B/Fiber~A ratio. This should not be the
case if Fiber A and B have simply a different transmission
coefficients, because they are both simultaneously observing
(almost) the same solar spectrum.

The NaI lines (reaching close to the zero level, see
Fig.~\ref{FBsFA}) are more sensitive to any unaccounted for, even
small, zero level shift. This is not the case for the weaker solar
lines.

Knowing that a zero level shift is present in the extracted
spectra, we search below for some plausible explanations.

\subsubsection{An instrumental zero level shift}

Over the SOPHIE CCDs, the lowest counts are not equal to zero
(even where no counts should be registered), as for instance at
the shortest wavelengths, where the instrument sensitivity drops
sharply. For this reason, we assumed the possibility of a zero
offset in the SOPHIE extracted counts. We thus subtracted a value
from the SOPHIE spectra to make the broad NaI solar lines wings
disappear (as they should). This correction is compatible with the
data, as reported in Table~\ref{table_fond}.

\begin{table}
\caption{SOPHIE fluxes as extracted from the data sets, both at
the minimum of deep absorption signatures and where the flux is
maximum. Times are for mid-exposures. Flux units are arbitrary but
comparable in relative terms because they correspond to the raw
data (CCD ADU) corrected for by the same pipe line. }
\begin{tabular}{cccc}
 \hline
 Fiber ; &  Min. flux  & Min. flux  &   Max. \\
 time (UT) & in NaI &  in O$_2$ & flux \\
 \hline
 \hline
A ; 20:44   &  0.0007   &  0.0005  &  0.0205 \\
 \hline
B ; 20:44  &  0.004  &  0.002  &  0.106 \\
 \hline
 \hline
A ; 02:56  &  0.0018      &  0.0010      &  0.0514 \\
 \hline
B ; 02:56  &  0.0016      &  0.0008      &  0.0471 \\
 \hline
 \hline
\end{tabular}
\label{table_fond}
\end{table}

We tried three zero shift level scenarios~:

\begin{enumerate}

\item we assumed that stray light due to the nearby full Moon is
present. However, subtracting various proportions of the full moon
spectrum to the other spectra did not induce any attenuation of
the NaI solar wings and on the contrary produced many additional
solar lines;

\item we supposed that a constant value has to be subtracted from
all data, proportional to the total flux gathered in each
spectrum. Again no satisfactory solution was found;

\item we assumed that a unique constant value had to be used for
all spectra, a correction only related to the detector and
independent of the total flux. Then we found that by subtracting
0.0003$\pm$0.0001 from all SOPHIE extracted values, it was
possible to completely wash out the NaI solar wings. This 0.0003
value represents a maximum relative continuum correction of 14\%
in the blue, decreasing to 1.5\% in the red, for the fiber A
spectrum. For fiber B, which is brighter, the relative continuum
correction is one order of magnitude lower.

\end{enumerate}

We verified that this last correction (3.) was also adapted at the
end of our study, because no broad wing solar NaI lines signatures
should show up in any information related uniquely to the Earth's
atmosphere. The corresponding final result (with and without this
correction) is shown in Sect.~\ref{discussion}, Fig.~\ref{NaI}.

This simple third correction (3.) does not seem to be
wavelength-dependent, because as we show in
Table~\ref{table_fond}, at the wavelengths of the very deep O$_2$
absorption lines near 6\,900~\AA , it is clear that this
correction could not be larger than the minimum flux seen at the
bottom of the absorption lines, and thus has to be somewhere in
the 0 to 0.0005 range, which we found to be the case. A
posteriori, we will also see that the final results are all
compatible with this correction.

After empirically evaluating the type of correction needed, we
found that another possible explanation of atmospheric origin may
be given and we develop it in the next section.

\subsubsection{An astrophysical zero level shift : the ring effect}

Indeed, following a suggestion of our referee, we recalled that
some of the sunlight going through the Earth atmosphere is
refracted in the lower layers of the atmosphere, producing for an
observer on the Moon within the Earth umbra, an emission ring all
along the Earth limb.

This ring effect can contribute here in two ways. First, it can
add a constant intensity (a few percent shift of the zero level)
to sunlight as seen in scattered light from the sky, to the extent
that the penumbral light contains not just transmitted light, but
also forward-scattered light~; this should slightly dilute the
penumbral spectrum and introduce a zero level correction of type
(2). Second, the light reflected from the lunar surface itself
will contain the ring effect, again a few percent additive
background spectrum, a correction of type (2.) or (3.).

   The ring relative intensity and spectral signature
were already observed during Lunar eclipses through the study of
the lunar light reflected from the umbral regions, only lit by the
Earth ring. Danjon (1936) very early understood that the umbral
luminosity was related to the transparency of the Earth lower
atmosphere and after observing several lunar eclipses, he proposed
a related five level scale for the lunar umbral luminosity.

Observation revealed (as in {\it e.g.} Hernitschek et al. 2008)
that a 5 magnitudes drop exists at the penumbra-umbra (PU)
transition, showing that the ring contribution to the penumbral
flux is on the order of 1~\%. Furthermore, the ring contribution
decreases when moving away from the PU transition deeper within
the umbra, as the expected signature of forward-scattered light.
The ring contribution within the penumbral region is indeed
proportional to the length of the Earth's limb ($L_E$) just in
front of the solar disk appearing on top of the Earth limb.

Thus, when moving in the penumbral domain away from the PU
transition, the direct solar contribution increases like the
surface of the solar disk above the Earth limb ($\sim$$L_E^2$, see
Fig.~\ref{geometry}), while the ring contribution increases as the
length of the Earth's limb covered by the solar disk
($\sim$$L_E$). This means that for our observations, the {\it
relative} ring effect contribution should be stronger in Fiber~A
than in Fiber~B according to the geometry of our 20h44 UT
observation, for which Fiber~A, presenting a lower signal, is
closer to the PU transition than Fiber~B. The ring effect
contribution in the reflected Lunar light seems thus to better
correspond to our empirical correction scenario (3.).

This is confirmed in the Gedzelman \& Vollmer (2008) simulation of
irradiance of lunar eclipses. They furthermore show that the ring
contribution at the PU transition should present an almost flat
spectral signature from 4000 to 7000\AA\ (see their Fig. 1), {\it
i.e.} over the whole SOPHIE spectral range. This was directly
observed during the 2008 August 16 lunar eclipse by Pall\'e et al
(2009, see top of their Fig.~S3 of the supplementary section), who
detected only weak spectral signatures deeper in the umbra as also
predicted by Gedzelman \& Vollmer (2008).

In summary the ring contribution should be on the order of 1~\%\
of the signal in Fiber A, and be at a relatively lower level in
Fiber B with no wavelength dependence. This is almost exactly what
we have found from our empirical study (3.), which presents the
same 0.0003 zero level shift in both Fibers, {\it i.e.} indeed on
the order of 1~\%\ in Fiber~A and relatively less in Fiber~B (see
Table~\ref{table_fond}). Additional discussions in
Sect.~\ref{discussion} will show that this correction is on the
same order near the NaI doublet lines at 5\,900~\AA\ and near the
molecular O$_2$ lines at 6\,900~\AA\ and thus is indeed weakly
variable with wavelength.

The ring effect could also explain why the ratios (Eclipse
A)/(Full Moon A), (or …B) could also be affected, ratios that we
are going to use and describe in the following sections.

Both these instrumental and ring effects are probably present in
our study and need to be corrected for empirically. In future
studies we will also try to observe umbral spectra in order to
have a more quantitative evaluation of the ring effect and thus
include its contribution in our analysis.

\section{Quantitative analysis}
\label{Q_analysis}

We present in this section the approach developed to extract a
quantitative information from the observed spectra related to the
altitude at which the Earth atmosphere becomes transparent as a
function of wavelength, a result very similar to the one obtained
for a transiting planet.

\subsection{Determination of the unabsorbed full Moon and eclipse
spectra}

These unabsorbed spectra are simply recomputed by dividing the
observed spectra $O_{a} (\lambda)$ by the transmission function $T
(\lambda)$ at the power of the precise air mass of the
corresponding observation~:

\begin{equation}
  I (\lambda) =  O_a (\lambda) / T^a (\lambda). \label{eqt100}
\end{equation}

All these calculated corrected spectra are those an observer
should have obtained from above the Earth atmosphere. We will note
these spectra $E (\lambda)$ and $F (\lambda)$, corresponding to
the ``eclipse'' and ``full Moon'' corrected spectra respectively.

\subsection{Earth's atmosphere effective thickness}

\begin{figure}[h]
   \centering
\resizebox{\columnwidth}{!}{\includegraphics[width=6cm]{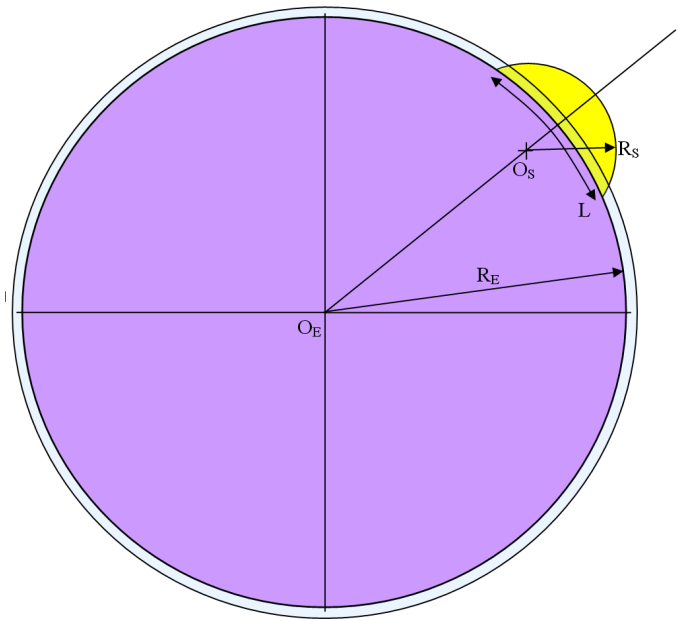}}
   \caption{Geometry of a crescent of Sun, $S$, above the Earth's reference limb
   (O$_S$ and $R_S$ are the center and radius
   of the Sun, O$_E$ and $R_E$ those of the Earth).
   The scene (not to scale) is observed from the Moon at the location where are positioned
   fibers A or B within the penumbra. From the observational knowledge
   of $S_A$ or $S_B$ as extracted from Eq.~\ref{eq200} (see text), the
   entire defined geometry of the observations allows a direct evaluation of the corresponding
   arc lengths, $L_A$ or $L_B$.}
\label{geometry}
\end{figure}

\begin{figure}[h]
   \centering
\resizebox{\columnwidth}{!}{\includegraphics[width=6cm]{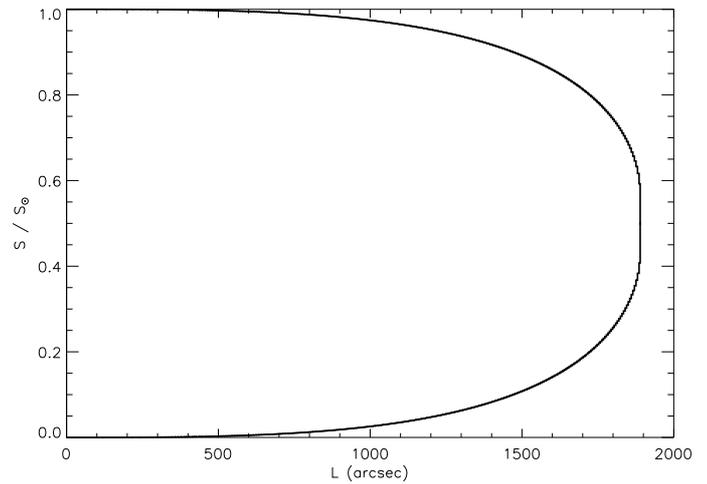}}
   \caption{$L = f(S)$ function computed here for the geometry of the eclipse:
   S is the surface of the solar crescent above the Earth's reference limb -
   as observed from the Moon at the location of either fiber A or B within the penumbra.
   From there, the diameters of the Earth and the Sun are 6\,768.0 and
   1\,890.6 arcsec respectively.
   This geometry implies the relation shown between the length of the Earth's limb covered by
   the Sun $L$, and the surface of the solar crescent, normalized to the
   total solar disk surface $S_\odot$ as seen from the Moon. Since $S_A / S_\odot$ or
   $S_B / S_\odot$
   are known from observations, the corresponding
   arcs lengths $L_A$ or $L_B$ are also known, and thus $h(\lambda )$ evaluated.} \label{LfS}
\end{figure}

From the deep penumbra, where the spectrograph's fibers are
located during the observation, a crescent of Sun is seen above
the Earth's limb. The Moon irradiance is thus a mix of direct Sun
light (i.e. unaffected by the Earth's atmosphere), and solar light
that passed through the atmosphere, mostly along an arc $L$ where
the solar disk intersects the Earth's limb (Fig.~\ref{geometry}).
This mix of direct and absorbed solar light essentially reproduces
what is recorded during a transit. Let us note $S_\odot$ as the
surface of the full solar disk, $S$ the fraction of solar surface
visible above Earth's geometrical limb during the eclipse, and $h$
is an equivalent height over which the atmosphere can be
considered opaque, at each wavelength, from the point of view of
the geometrical cross section.

 The total column density along a grazing line of sight passing through the
terminator at different altitudes could be estimated following the
Fortney et al. (2005) formulation: the optical depth, $\tau$, in a
line of sight grazing the Earth limb at an altitude $h$ is given
by

\begin{equation}
\tau(\lambda,h)\approx \sigma(\lambda)n(h)\sqrt{2\pi R_{E} H},
\label{tau_lambda}
\end{equation}

where $R_{E}$ is the Earth radius, $H$ the atmosphere scale
height, and $n(h)=n_{(h=0)} exp(-h/H)$ the volume density at the
altitude $h$ of the main absorbent with a cross section
$\sigma(\lambda)$. The scale height is given by the relation
$H=kT/\mu g$, where $k$ is the Boltzmann constant, $T$ the
temperature, $\mu$ the mean mass of atmospheric molecules times
the mass of the proton, and $g$ the gravity at the Earth radius,
$g = M_{E}{\rm G}/R_{E}^2$ with $M_{E}$ equal to the Earth mass
and G the gravitational constant.

For any wavelength, the line of sight becomes opaque at an
effective altitude $h$ for an effective optical thickness of
$\tau_{eff}$~=~0.56 as demonstrated to be appropriate for most
atmospheric conditions by Lecavelier des Etangs et al. (2008a). We
will then be able to compare our equivalent effective altitude $h$
as a function of wavelength to model calculations of effective
altitudes.

The effective altitude of the atmosphere at a wavelength $\lambda$
is calculated by solving the equation
$\tau(\lambda,h)=\tau_{eff}$. Using the quantities defined above,
the effective altitude $h$ is given by Eq.\ref{z_lambda}.

\begin{equation}
h(\lambda)=H\ln \left((n_{(h=0)} \sigma(\lambda) / \tau_{eff})
 \sqrt{2\pi  R_E/kT \mu g}\right).
\label{z_lambda}
\end{equation}

In principle, a simple relation links the two $E (\lambda)$ and $F
(\lambda)$ corrected spectra due to the geometry of the problem

\begin{equation}
  E (\lambda) =  F (\lambda) \times  \frac{S - L \times h  (\lambda)}{S_\odot}.  \label{eq1}
\end{equation}

\subsection{Evaluation of $h (\lambda)$, the effective height of the Earth atmosphere.}

\begin{figure}[h]
   \centering
\resizebox{\columnwidth}{!}{\includegraphics[width=6cm]{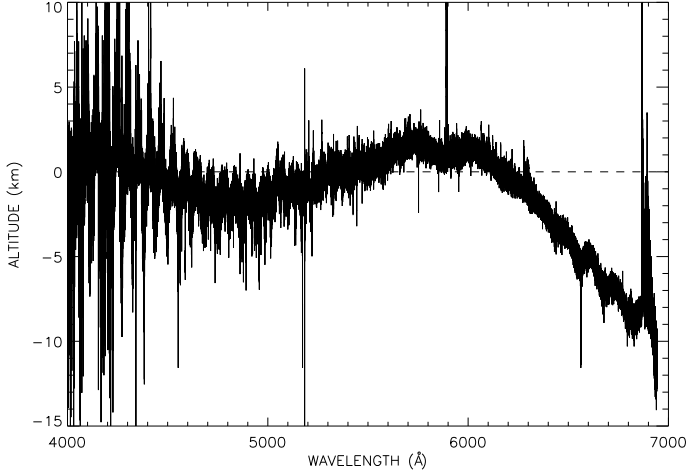}}
   \caption{Absorbing atmosphere thickness versus wavelength,
   evaluated according to Eq.~\ref{ratioAB}
   for the 20h44 UT eclipse observation associated to the 02h56~UT full Moon ones.
   The reference altitude equal to 0~km (dashed line) has been chosen
   for $\lambda$~=~4\,530~\AA .
   The profile is dominated by the Chappuis band of ozone from 4\,800 to 6\,900~\AA,
   while some signatures due to
   sodium and oxygen show up near 5\,890~\AA\ and 6\,900~\AA\ respectively.
   In the blue region (4\,000 to 4\,800~\AA), the
   Rayleigh scattering is clearly seen. Across that region, the signal drops and produces more
   noise over the extracted altitudes, also revealing there that the SOPHIE orders are not
   fully corrected. This is also true longward from about 6\,500~\AA .} \label{figure_h}
\end{figure}

\begin{figure}[h]
   \centering
\resizebox{\columnwidth}{!}{\includegraphics[width=6cm]{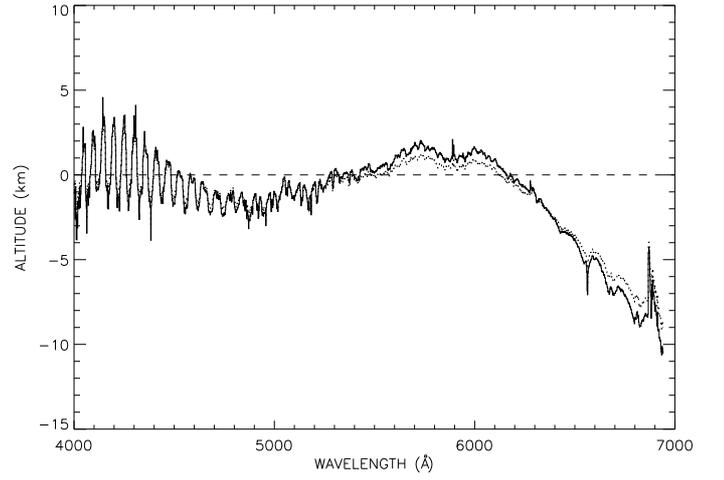}}
   \caption{Same as Fig.~\ref{figure_h}, but here the $h(\lambda)$
   variations are binned over 200 pixels ($\sim$2~\AA ) for clarity.
   Here the altitude variations are, following Eq.~\ref{ratioAB}, extracted from the
   eclipse observations of 20h44 UT with either the 02h56~UT (solid line) or the
   00h12~UT (dotted line) full Moon ones. Without the binning the
   difference was not visible. Here we can note that both possible perturbations
   related to either the BERV shift as well as the air mass correction
   are indeed negligible (see text). On the contrary some broadband
   variations could be seen, probably linked to albedo
   variations of the different lunar regions where the fibers were
   pointed (see Table~\ref{table_position}). In the blue region
   (4\,000 to 5\,000~\AA), the fringes are due to an imperfect
   correction of the SOPHIE orders and not to classical CCD interferential fringes (see text).}
   \label{two_h}
\end{figure}

We can define a reference altitude $h(\lambda_0)$~=~0 at an
arbitrary wavelength $\lambda_0$. Then, $S$ is more precisely the
fraction of solar surface visible above the Earth's limb
corresponding to our reference altitude $h(\lambda_0)$. It follows
that Eq.~\ref{eq1} reduces to

\begin{equation}
  E (\lambda_0) =  F (\lambda_0) \times  \frac{S}{S_\odot}. \label{eq200}
\end{equation}

The $S / S_\odot$ ratio can be geometrically calculated for a
given observation time, but may also be directly estimated from
the $E (\lambda_0)$ and $F (\lambda_0)$ values of the flux
measured during and outside the eclipse at $\lambda_0$. We decided
to take the reference wavelength over a 20~\AA\ range centered at
$\lambda_0$~=~4\,530~\AA , where the absorption in the Earth
atmosphere (see Fig.~\ref{transmission}) is mainly due to Rayleigh
scattering. Because the altitude above sea level (corresponding to
the same reference wavelength) is the standard outcome from any
atmospheric model calculation (see next section), it will be then
simple to shift all our evaluated altitudes relative to our
reference altitude $h (\lambda_0)$ into ``real'' altitudes, above
sea level. Indeed all our altitude evaluations $h (\lambda)$
relative to $h (\lambda_0)$, could be positive or negative
depending on more or less absorption at $\lambda$ than at
$\lambda_0$.

The ratio of Eqs.~\ref{eq1} and \ref{eq200} gives for Fiber~A

\begin{equation}
  \frac{E_A (\lambda)}{E_A (\lambda_0)} = \frac{F_A
  (\lambda)}{F_A (\lambda_0)} \times [1 - \frac{L_A}{S_A} \times h (\lambda)],
  \label{ratioA}
\end{equation}

and for Fiber~B

\begin{equation}
  \frac{E_B (\lambda)}{E_B (\lambda_0)} = \frac{F_B
  (\lambda)}{F_B (\lambda_0)} \times [1 - \frac{L_B}{S_B} \times h (\lambda)],
  \label{ratioB}
\end{equation}
in which $L_A$ and $L_B$ are the lengths of the arc of limb at the
$h(\lambda_0)$ level as seen from the locations of Fiber~A and
Fiber~B respectively, as projected on the Moon, and similarly
$S_A$ and $S_B$ are the corresponding surfaces of the Sun above
these Earth's limbs.


The difference between the ratios for Fibers A and B,
~(Eq.~\ref{ratioA} -- Eq.~\ref{ratioB}), is

$$  \left( \frac{E_A (\lambda)}{E_A (\lambda_0)} \times \frac{F_A
(\lambda_0)}{F_A
  (\lambda)} - \frac{E_B (\lambda)}{E_B
  (\lambda_0)} \times \frac{F_B (\lambda_0)}{F_B (\lambda)} \right) =
$$

\begin{equation}
 ~~~~~~~~~~~~~~~~~~~~~~~~ = \left( \frac{L_B}{S_B} - \frac{L_A}{S_A} \right)
  \times h (\lambda).
\label{ratioAB}
\end{equation}

Therefore, the measurements (left side of Eq.~\ref{ratioAB}) are
proportional to $h$.

However, to fully solve the problem, one needs to evaluate the
factor of proportionality in front of $h$. We know the relative
angular diameters of both the Earth (above the Earth's reference
limb) and the Sun, as seen from the Moon at the time of the
observations, which unambiguously defines (geometrically) how far
above that selected Earth's limb reference the solar disc emerges.
From a simple geometrical calculation (see Fig.~\ref{geometry}),
one can evaluate the corresponding arc lengths $L_A$ and $L_B$,
which are directly derived from the relation $L = f(S)$ that we
have calculated for the time of the observations, as shown in
Fig.~\ref{LfS}. Finally, the direct measurements of $S_A$ and
$S_B$ are extracted from the observations through Eq.~\ref{eq200}.

For example, from our best observations made on August 16 at
20h44~UT during the eclipse, compared to the full Moon
observations made on August 17 at 02h56~UT, we find the following
corrected flux ratios in the 4\,520~-~4\,540~\AA\ reference
spectral band~:

$E_A (\lambda_0)/F_A (\lambda_0)$~=~0.44~\%\, a value similar to
$S_A / S_\odot$ as seen from the Moon where Fiber~A is located;
this leads to $L_A = 569$~arcsec;

$E_B (\lambda_0)/F_B (\lambda_0)$~=~3.12~\%\, a value similar to
$S_B / S_\odot$ as seen from the Moon where Fiber~B is located;
this leads to $L_B = 1\,063$~arcsec.

In conclusion, from all parameters fully known at a given
observing time, Eq.~\ref{ratioAB} gives access to direct
evaluations of $h (\lambda)$, either in arcsec seen from the Moon
or in km above (or below) the Earth's reference limb at $h
(\lambda_0)$.

The extracted values of $h (\lambda)$ are shown in
Fig.~\ref{figure_h}, obtained when using the 20h44~UT eclipse
observations compared to the 02h56~UT full Moon ones at about the
same air mass and over the same regions on the Moon. The extracted
altitudes are relative to the reference altitude $h (\lambda_0)$,
here equal to~0. As expected, the evaluations can be either
positive (higher altitudes) or negative (lower altitudes).

To check the validity of several of our assumptions we calculated
the same $h (\lambda)$ function extracted again from the 20h44~UT
eclipse observations compared to both full Moon observations, the
one at 02h56~UT as well as the 00h12~UT one made at different air
masses, BERV and over distant regions on the Moon (see
Table~\ref{table_position}). This allows us to check that our
evaluations are not sensitive to air masses or BERV variations
(see Fig.~\ref{two_h}), because both extracted $h(\lambda)$
functions are very similar in all their details. Although nearly
identical (non discernable at full spectral resolution as in
Fig.~\ref{figure_h}), the result indicates however that some broad
band variation is present. It is not caused by an improper air
mass correction because the produced effect cancels in two
spectral regions: the first one at 4\,530~\AA\ (by definition,
since this is our reference wavelength) and again around
6\,300~\AA . That the observed relative variation of the
transmission remains small in the blue and more important in the
red is indeed in contradiction with an air mass correction, which
has to be stronger in the blue (see Fig.~\ref{transmission}). We
thus conclude that the observed relative variation is more
probably due to lunar albedo changes over different lunar regions.

This broadband effect could however be the cause of additional
systematic errors in the altitude evaluations.

\subsection{Error estimation}

A simple visual analysis of Figs.~\ref{figure_h} and \ref{two_h}
gives some idea of the errors estimated for the $h(\lambda)$
evaluations. First in Fig.~\ref{figure_h}, the error due to the
photon noise in the observations is $\sim$~$\pm$1~km for all data
points shown but clearly increases toward shorter wavelengths
(below 5\,000~\AA ) due to the decreasing sensitivity of the
instrument toward these wavelengths. Furthermore in this spectral
region the impact of the improper order corrections is more and
more important as more easily visible in Fig.~\ref{two_h}. This
effect induces an additional systematic error below 5\,000~\AA\
(of $\sim$~$\pm$2.5~km) and above 6\,400~\AA\ (of
$\sim$~$\pm$1~km). Finally, although not all possible lunar
terrains were sampled, we can guess from the different regions
observed that an additional systematic error due to lunar albedo
variations could be on the order of $\sim$~$\pm$1~km. This means
that for our first attempt to estimate the Earth $h(\lambda)$
function we cannot produce absolute altitudes to better than
approximately $\sim$~$\pm$2.5~km.

\section{Discussion}
\label{discussion}

\subsection{Model calculation}

The information extracted from the presented spectra is
$h(\lambda)$, thus we are exactly in the situation of an
extrasolar planetary transit. In effect, a spectrum recorded
during a transit is \textit{not} the spectrum of the absorbing
species, but only the spectral signatures (in terms of altitude)
of the \textit{highest} absorbing species within the observed
atmosphere. For instance, in the central region of the spectrum
(Fig.~\ref{figure_h}) the main feature is the Chappuis band of
ozone, corresponding to altitudes known to be in the 30~km range,
other detectable species are only those able to efficiently absorb
above. As we will see, oxygen and sodium are also detected with
the present observations, but not H$_2$O.

In order to properly interpret our observations, we have to use
model calculations to estimate the altitude at which the Earth
atmosphere becomes transparent, as a function of wavelength and
for a grazing line of sight exactly as in a transit situation.

Standard Earth atmospheric model calculations are from {\emph
e.g.} Ehrenreich et~al. (2006) or Kaltenegger \&\ Traub (2009).
From these model calculations, one can evaluate the effective
Earth radius at each wavelength, and from it, the effective height
of the Earth atmosphere as a function of wavelength. From Fig.~3
of Kaltenegger \&\ Traub (2009), the effective altitude
corresponding to the tip of the Chappuis band absorption (near
6\,000~\AA ), is on the order of 30~km, revealing that the
atmosphere becomes nearly opaque below this altitude at this
wavelength.

\begin{figure*}[!ht]
   \centering
\resizebox{\textwidth}{!}{\includegraphics[width=13cm]{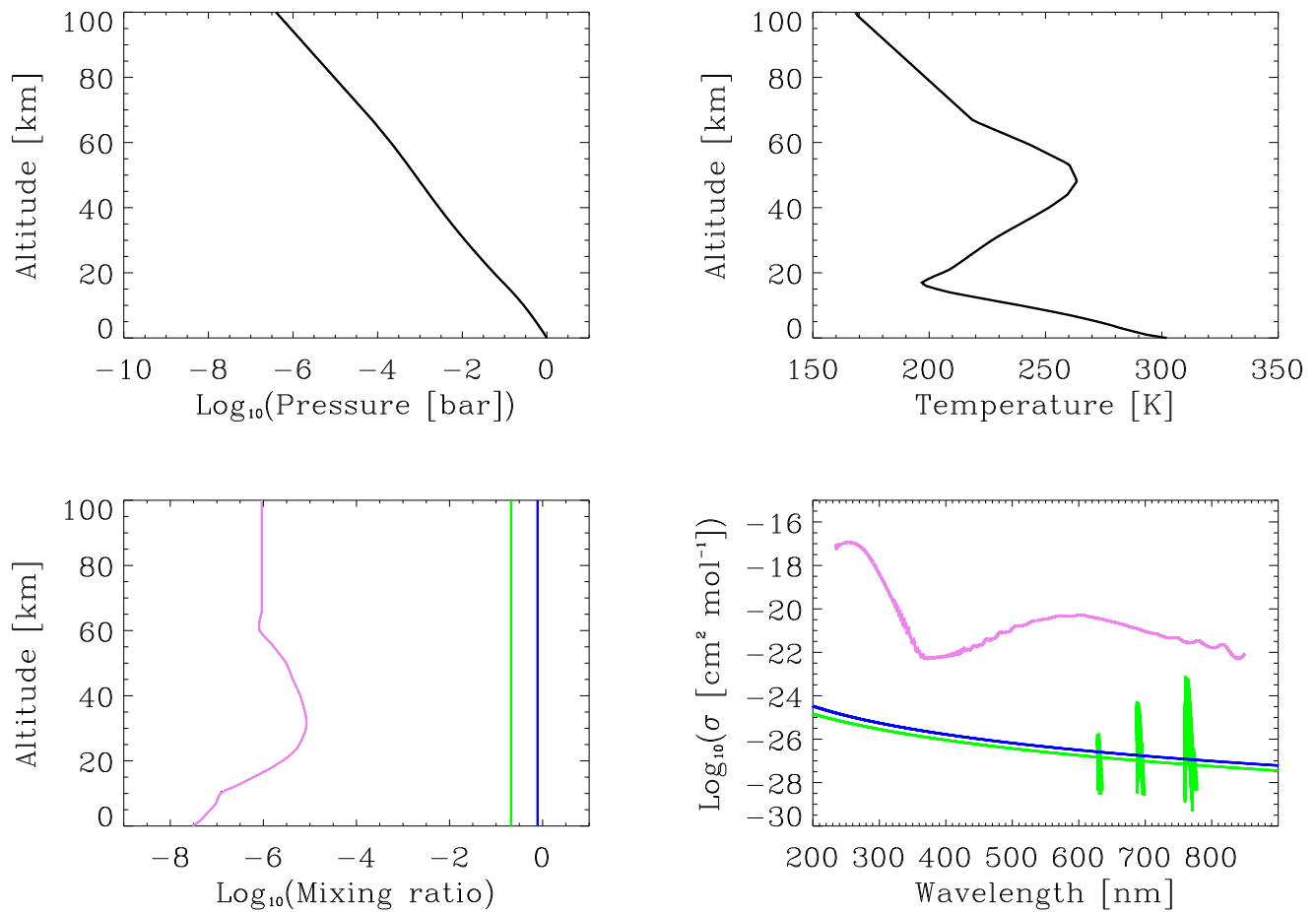}}
   \caption{``Standard'' atmosphere used in the to model calculation
   (see text)~: {\bf Upper left} the pressure profile, {\bf upper right}
   the temperature profile, {\bf lower left} the N$_2$ (blue), O$_2$ (green)
   and O$_3$ (pink) mixing ratio
   as a function of altitude (km)
   and {\bf lower right} the N$_2$ (blue) Rayleigh, O$_2$ (green)
   Rayleigh and molecular bands and O$_3$ (pink) cross sections as a function
   of wavelength (in nm).} \label{atmoprofiles}
\end{figure*}

In our model calculation we used an unidimensional single
scattering transmission model based on the model described by
Ehrenreich et~al. (2006) to obtain the theoretical transmission
spectrum of the terrestrial atmosphere. The pressure, temperature,
and mixing ratio profiles of the considered atmospheric components
are shown in Fig.~\ref{atmoprofiles} and come from the US Standard
Atmosphere 1976 spring-fall pressure-temperature profile
(\cite{cosea1976}; \cite{cox2000}). For simplicity, we chose to
only include molecular nitrogen (N$_2$), molecular oxygen (O$_2$),
and ozone (O$_3$) in the model. Molecular nitrogen contributes to
the transmission spectrum via the Rayleigh-scattering of light.
The Rayleigh-scattering cross section of N$_2$ depends on the
N$_2$ refractive index, which is calculated according to Sneep \&
Ubachs (2005). Within the SOPHIE spectral range, molecular oxygen
contributes to the transmission spectrum through
i)~photoabsorptions by the forbidden
$^1\Sigma_g^+$--$^3\Sigma_g^-$ transition bands B (1--0) at
6\,880~\AA\ and $\gamma$ (2--0) at 6\,280~\AA, and ii)~through
Rayleigh scattering. The line parameters for the O$_2$ bands are
extracted from the HITRAN 2008 molecular spectroscopic database
(\cite{rothman2009}), scaled to the atmospheric temperature and
pressure profiles used, and convolved with a Gaussian profile with
full-width at half-maximum matching the spectrograph resolution.
The Rayleigh-scattering cross sections of O$_2$ are calculated
following Sneep \& Ubachs (2005) using refractive indices from
Bates (1984). We retrieved the UV/visible photoabsorption cross
section of O$_3$ at 293 K and 1 bar from the GEISA 1997 data base
(\cite{jacquinet1999}).

We recall here that we are not doing a fit of our data but just
compute altitude signatures related to a ``standard'' Earth
atmospheric model.

\subsection{Broadband signatures}

\begin{figure*}[!ht]
   \centering
\resizebox{\textwidth}{!}{\includegraphics[width=13cm]{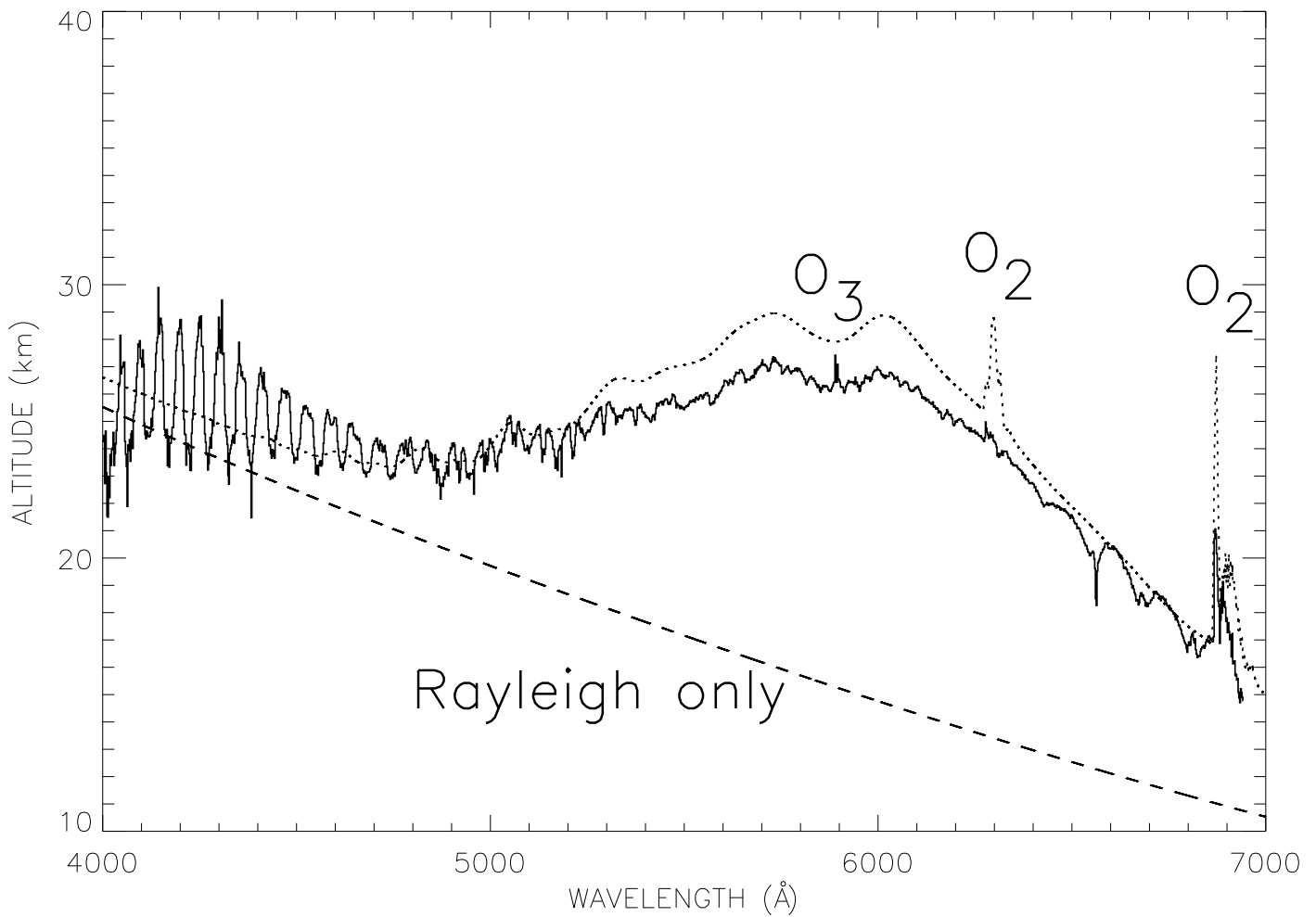}}
   \caption{Observed binned variations (solid line) of the effective altitude
   $h$ compared to model calculations (also binned).
   The Rayleigh-alone model calculation is shown (dashed line) along
   with the complete atmospheric model calculation (dotted line),
   which includes N$_2$, O$_2$ and O$_3$ (see text).} \label{h_model}
\end{figure*}

The variable effective altitude $h$ (as a function of wavelength)
is shown on Fig.~\ref{h_model} for both model calculations and as
extracted from the observations, the overall contribution of
O$_3$, O$_2$ and Rayleigh scattering due to both N$_2$ and O$_2$,
along with the Rayleigh contribution alone are also shown.

In the selected reference domain from 4\,520 to 4\,540~\AA , the
model calculation predicts an altitude equal to 23.8~km. Because
our observational evaluation of $h$ is {\it relative} to the
altitude in the reference domain, we shift our {\it relative}
$h(\lambda )$ evaluations by $+23.8$~km to be comparable to the
model calculation. Since furthermore our reference domain falls at
the tip of a SOPHIE echelle order fluctuation, we corrected by an
additional systematic shift of about 1.5~km (half of the echelle
fluctuation). The model and observations are shown in
Fig.~\ref{h_model}, where a 25.3~km reference altitude is used in
the 4\,520 to 4\,540~\AA\ reference domain range.

Our observed evaluations as shown in Fig.~\ref{h_model} are binned
over 200 pixels ($\sim$2~\AA ) to better underline the observed
broadband variations. It can be seen that the errors over the
broadband signatures are mainly due to systematic errors, as
opposed to photon noise, and are on the order of the SOPHIE
approximate order correction, {\it i.e.} $\sim \pm 2.5$~km.

The comparison between the observed variations and model
calculations is very satisfactory. Even the observed ozone
distribution seems probably less extended in altitude than the
standard model by about 2~km, a very plausible situation because
ozone variations are known to exist around the Earth in terms of
both location and epoch (Millier et~al. 1979, Thomason \&\ Taha
2003, Borchi \&\ Pommereau 2007). There is no reason that the
selected standard atmospheric model exactly corresponds to the
Earth atmospheric conditions at the precise location over the
observed Earth's limb.

\begin{figure*}[!ht]
   \centering
{\includegraphics[width=13cm]{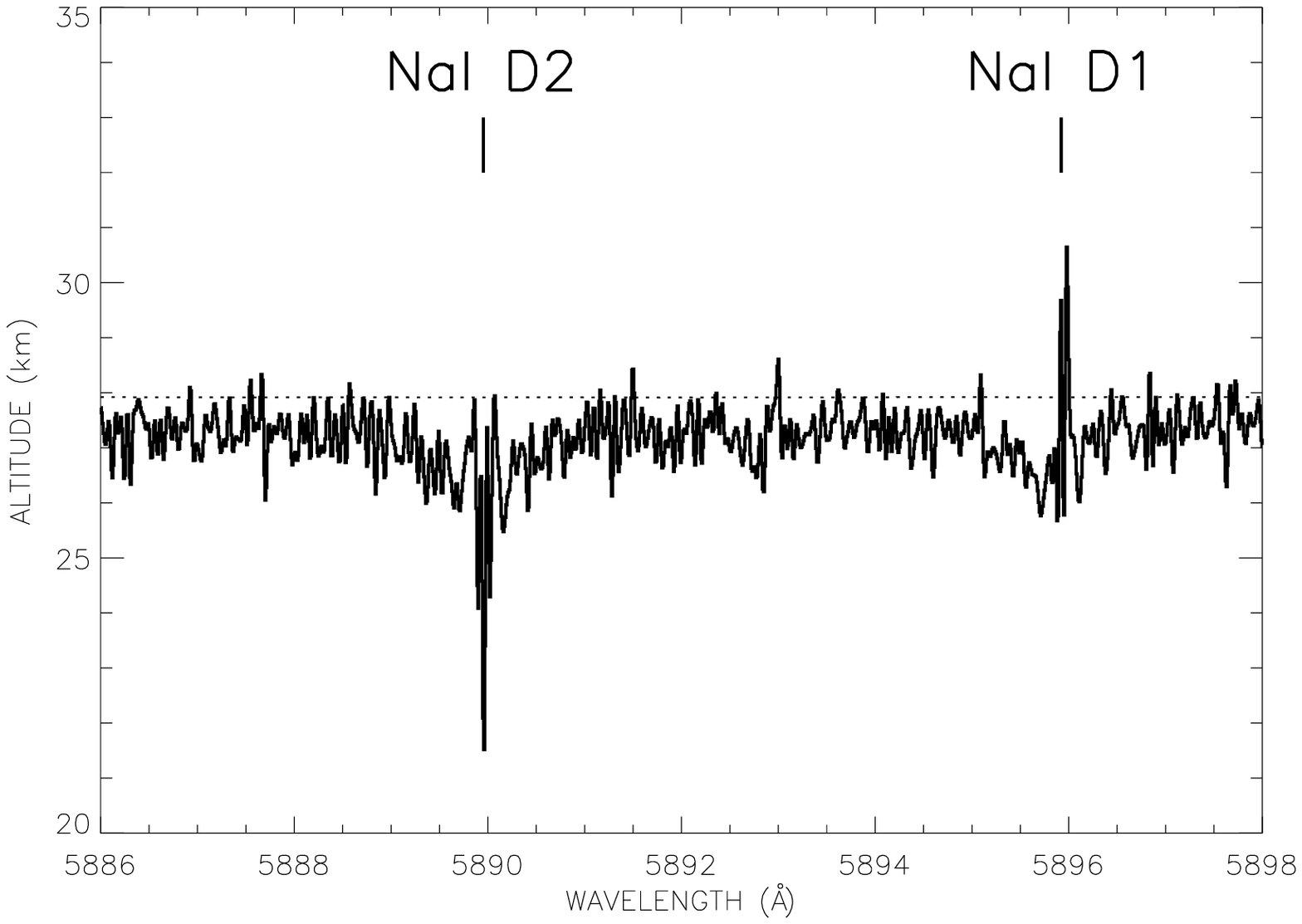}}
{\includegraphics[width=13cm]{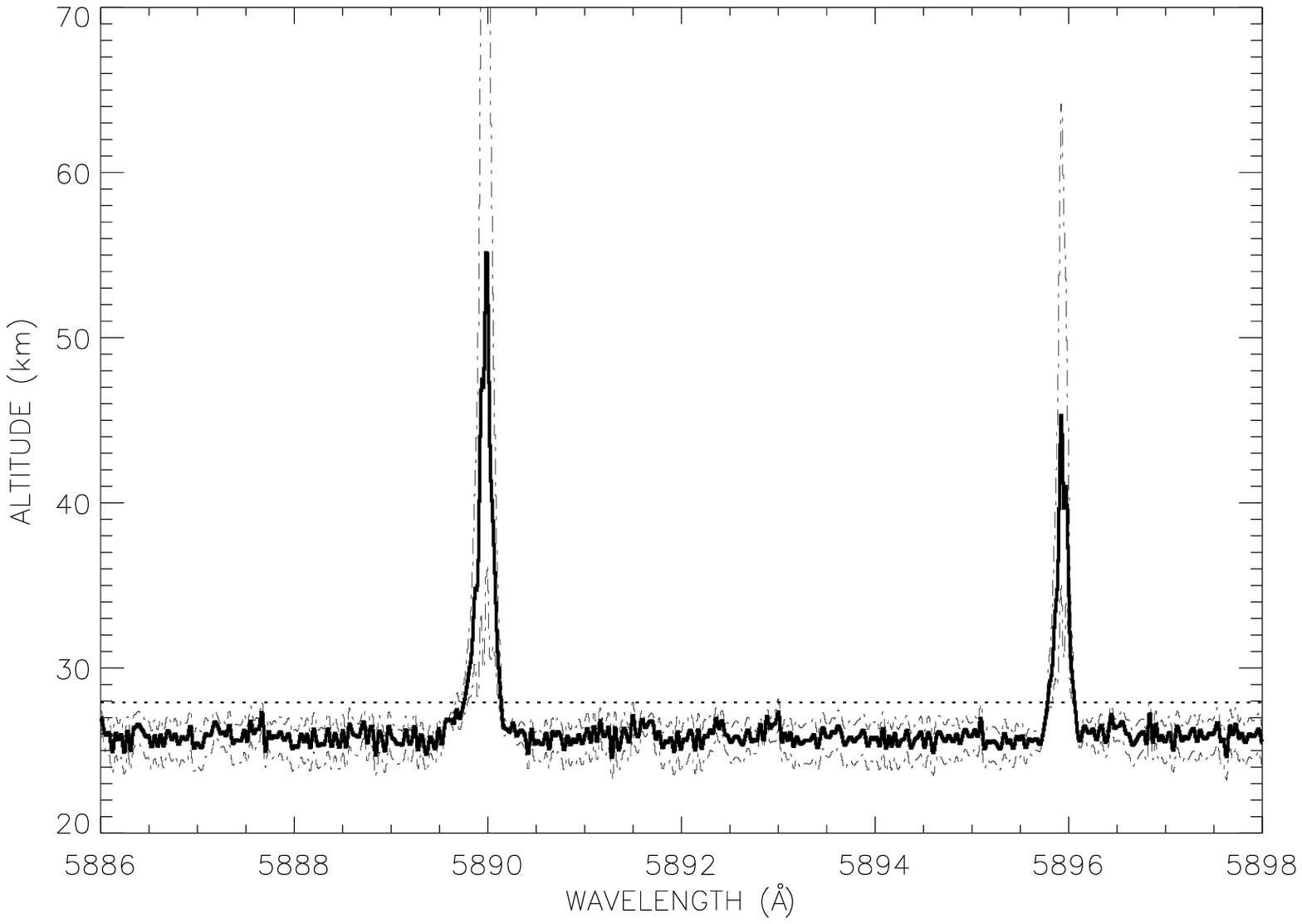}}
   \caption{ NaI doublet spectral region. The shifted $h(\lambda )$ variations
   (solid lines) are shown along with the level of the model ozone prediction (dotted line).
   {\bf Upper plot.} The zero level correction as detailed in Sect.~\ref{data_analysis}
   is not applied. The two NaI narrow atmospheric signatures do not show up while through
   our complete data analysis method the broad wings of the sodium solar lines are
   obviously still present, contrary to expectation (the solar broad lines wings
   should not appear in any information uniquely related to the Earth atmosphere). {\bf Lower plot.} The zero level
   correction by subtracting a 0.0003 value is now applied (see text).
   The NaI solar wings have disappeared, while the narrow NaI atmospheric signatures
   clearly show up. To show the impact of the zero level correction, the two
   evaluations corresponding to the 0.0002 and 0.0004 zero
   corrections are overplotted (thin dash-dotted lines).
   This gives a direct idea of the induced error bars on our evaluation.} \label{NaI}
\end{figure*}

\subsection{Narrowband spectral signatures}

The acute need for high spectral resolution is highlighted by the
detection of other species, which are producing signatures at
altitudes above the ozone level.

\subsubsection{Sodium detection}

The spectral region containing the NaI doublet is shown in
Fig.~\ref{NaI}. To repeat the discussion presented in
Sect.~\ref{data_analysis}, concerning the zero flux level
correction, the result without any correction applied is also
shown. Obviously, the atmospheric NaI doublet does not show up
without that correction, and furthermore the broad wings of the
two NaI solar lines are still present while our extraction process
should have entirely washed out any solar line signature, as it is
indeed almost perfectly the case in Figs.~\ref{figure_h} or
\ref{h_model}.

This correction introduces an additional systematic error, which
produces an altitude shift of about 1.5~km per 0.0001 step change
in the zero level (see Fig.~\ref{NaI}). This is small compared to
the systematic error due to the instrumental order correction (on
the order of $\pm$2.5~km), which thus remains the major cause of
uncertainty on our extracted altitude level.

Note here a very interesting point. We know that both H$_2$O and
O$_2$ produce strong absorption signatures in the Earth atmosphere
and, as shown in Fig.~\ref{transmission}, they have relatively
similar signatures near 5\,900~\AA\ (in the NaI lines vicinity)
for H$_2$O or around 6\,300~\AA\ for O$_2$. We will see in the
sections below that O$_2$ is indeed easily detected in the grazing
(transit like) observations of the atmosphere, while we see here
that none of the numerous H$_2$O lines present in that spectral
range (see Fig.~\ref{FBsFA}) are detected. The reason is very
simple: H$_2$O is hidden in the transit like observations because
it is at lower altitudes than ozone, while O$_2$, a major
constituent of the atmosphere, is still present in sufficient
quantities at higher altitudes, and indeed shows up above the
ozone layer. This is exactly what models do predict (see {\it
e.g.} \cite{kaltenegger2009}) and thus reveals the quality and
consistency of our analysis, from which only species thought to
show up are indeed detected.

The zero level correction induces additional very large errors on
the estimated altitude of the NaI spikes: the NaI 5\,890~\AA\ line
peaks up from 35 to 100~km, while the 5\,896~\AA\ one from 35 to
65~km altitudes.

NaI is known to be present in the Earth atmosphere within a layer
at about 92~km, presenting an average thickness of about 11~km
(\cite{moussaoui2010}). If it was completely opaque in the two NaI
lines, we should have found the same level in both lines at lower
altitudes than the 92~km, depending on the ratio between the
atmospheric line width ($\sim$0.03~\AA ) and the instrument
spectral resolution ($\sim$0.09~\AA ). This should lead to reduced
altitudes of about one third of 92~km above the 30~km ozone
signature, {\it i.e.} $\sim$50~km.

This could correspond to our 0.0002 zero level correction, leading
to both line levels at about 35~km altitude, a little low when
compared to our evaluation.

However, we can also directly evaluate the line's opacity in the
92~km altitude layer by using the observed average vertical sodium
column density of the layer to be on the order of
4.$10^{13}$~m$^{-2}$ (\cite{moussaoui2010}). Using the average
altitude, thickness, and ``vertical'' column density of the layer,
we can translate these numbers into an average ``horizontal''
column density at 92~km of altitude and found it to be on the
order of 3.$10^{11}$~cm$^{-2}$, leading to an optical thickness in
the core of the line on the order of 3 in the strongest NaI line
and 1.5 in the weaker one. Both lines are thus opaque, but not
strongly opaque. The known variability of the NaI layer could be
the reason for this slightly less opacity, which renders it
probable that a marginally opaque layer was observed during the
eclipse, then leading to two different altitudes for the observed
doublet spikes corresponding to the also plausible values as the
ones observed with our 0.0003 zero corrections, which are at 55
and 45~km respectively.

From this brief quantitative discussion, we note that the 0.0004
zero correction is certainly a limit because, in that case, the
altitude of the strongest line is at more than 100~km, {\it i.e.}
too high, even above the altitude of the layer itself.

The NaI atmospheric detection is certainly compatible with what is
known about the Earth layer, both qualitatively and
quantitatively. This shows that our high-resolution approach could
give us access to narrow line signatures, and thus gives us
confidence in searching for other species. We detail this search
in the following sections.

\subsubsection{The 6\,880~\AA\ oxygen molecular band}

Two molecular oxygen absorption bands have their signatures in the
observed spectral range. We used for them the same zero level
correction as for the NaI lines.

\begin{figure*}[!ht]
   \centering
{\includegraphics[width=13cm]{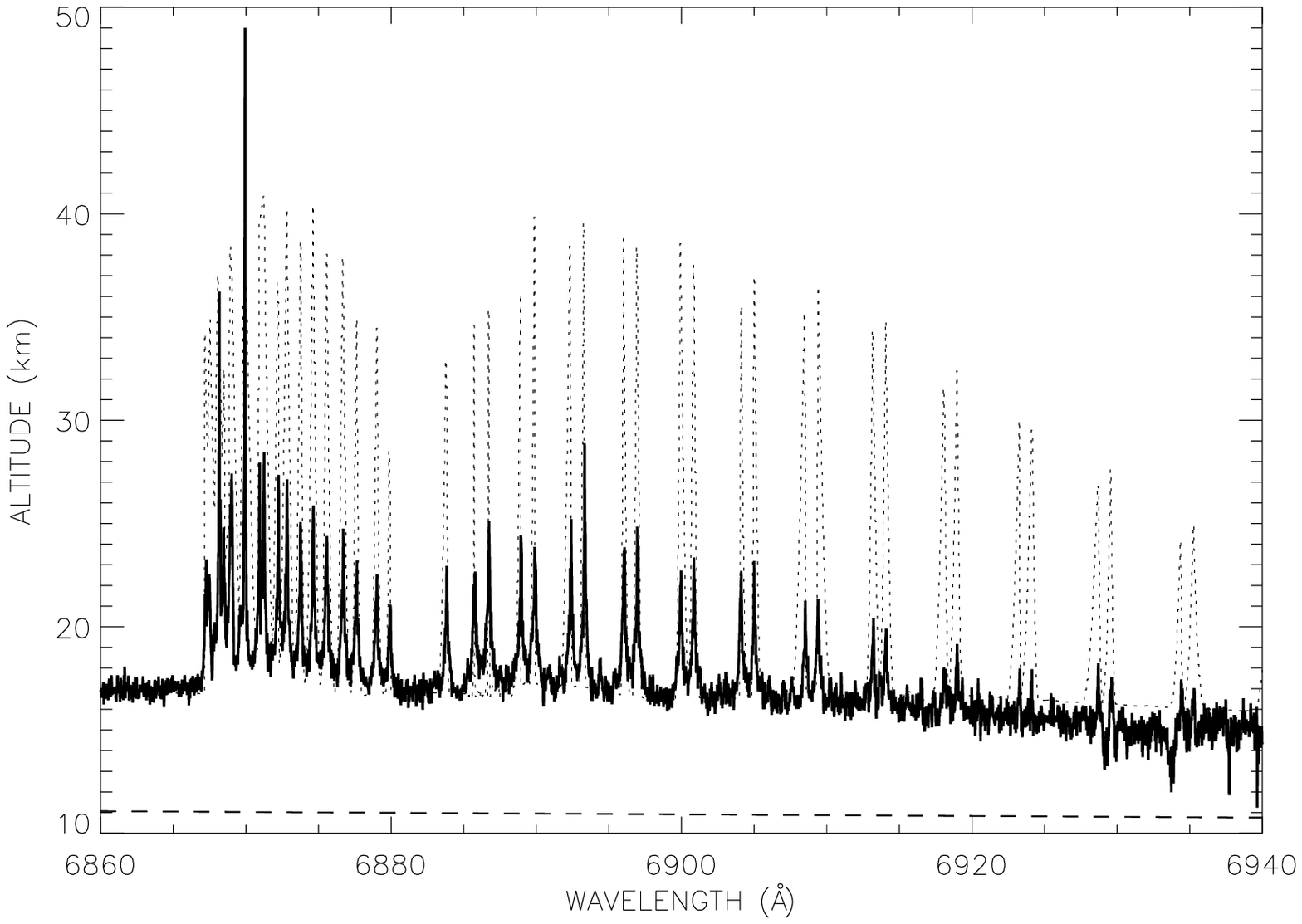}}
{\includegraphics[width=13cm]{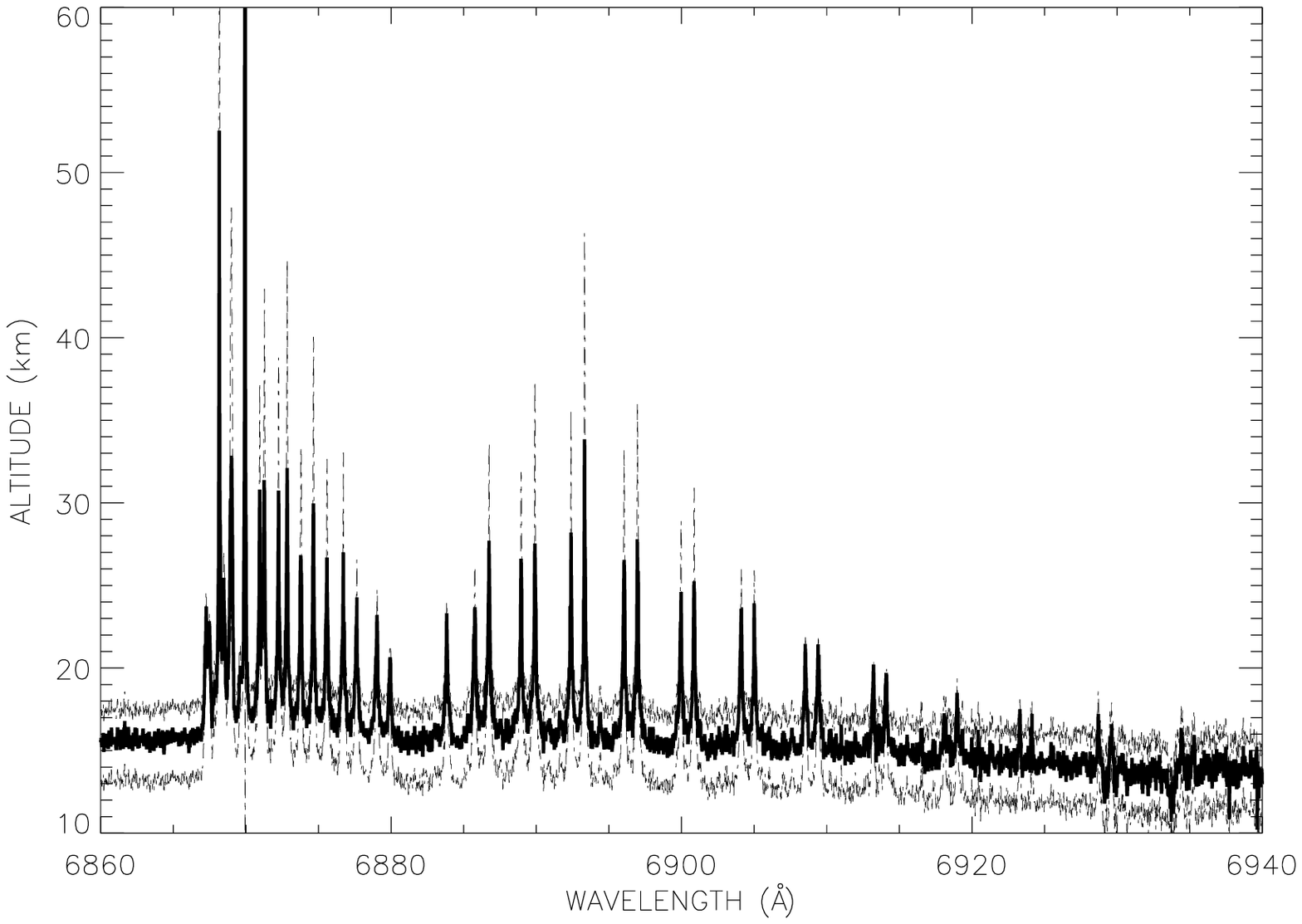}}
   \caption{O$_2$ forbidden $^1\Sigma_g^+$--$^3\Sigma_g^-$
transition band B (1--0) at 6\,880~\AA . The same background
correction as in Fig.~\ref{NaI} has been applied
   (see text). {\bf Upper plot.} The solid line represents the altitude information
   extracted from the {\it SOPHIE} data, while the dotted line is the model
   calculation, in which O$_2$ absorption is included
   from 6\,856~\AA\ upwards. The narrow atmospheric signatures due to
   the O$_2$ absorption are clearly seen.
   The sharp peaks perfectly match the model with respect to spectral
   positions, while their
   heights could be quite different.
   {\bf Lower plot.}
   The thick solid line again represents the evaluated altitude for the 0.0003 zero level
   correction, while the thin dash-dotted lines show the
   extreme possible variations of the estimated levels due to the 0.0002 (lower evaluations
   in the peaks and higher ones in the continuum)
   and 0.0004 (higher evaluations in the peaks and lower ones in the continuum)
   zero level corrections.}
   \label{O2_6900-0.0003}
\end{figure*}

The strongest one, the O$_2$ forbidden
$^1\Sigma_g^+$--$^3\Sigma_g^-$ transition band B (1--0) at
6\,880~\AA , is shown in Fig.~\ref{O2_6900-0.0003}. All molecular
band peaks are clearly present, undoubtedly signing the O$_2$
detection. To evaluate the impact of the zero level adjustment,
the three evaluations of $h(\lambda )$ made with the three values,
0.0003$\pm$0.0001, are also shown in Fig.~\ref{O2_6900-0.0003}.
This directly shows that in this spectral range, the zero level
adjustment adds a systematic error of about $\sim$$\pm$2.5~km in
the continuum and up to $\sim$$\pm$10~km in the peaks, showing
that their absolute estimates from these observations are quite
imprecise.

However, the average peak altitude and its relative height
variations are certainly more meaningful. In particular, the
average height of the 0.0004 correction produces signatures closer
to the model predictions, suggesting that the zero level
correction may be slightly wavelength-dependant or that its
acceptable range is reduced to 0.00035$\pm$0.00005. One can also
note that the observed peak heights when moving redward seem to
drop more rapidly than in the model prediction, an obvious
indication of a cooler atmosphere than the ``standard'' one
selected for the model calculation. This suggests that better
adjusted Earth atmospheric model parameters in terms of vertical
temperature-pressure profile are needed. This is however beyond
the scope of this paper, in which we just intend to show that some
species detection is feasible according to that observational
approach in both qualitative and quantitative terms. The fine
adjustment of model parameters will be done in future papers
according to future observations to be made, all along a complete
lunar eclipse to search for Earth atmospheric variations along the
Earth limb sampled during the eclipse evolution.

\subsubsection{The 6\,280~\AA\ oxygen molecular band}

\begin{figure}[h]
   \centering
\resizebox{\columnwidth}{!}{\includegraphics[width=6cm]{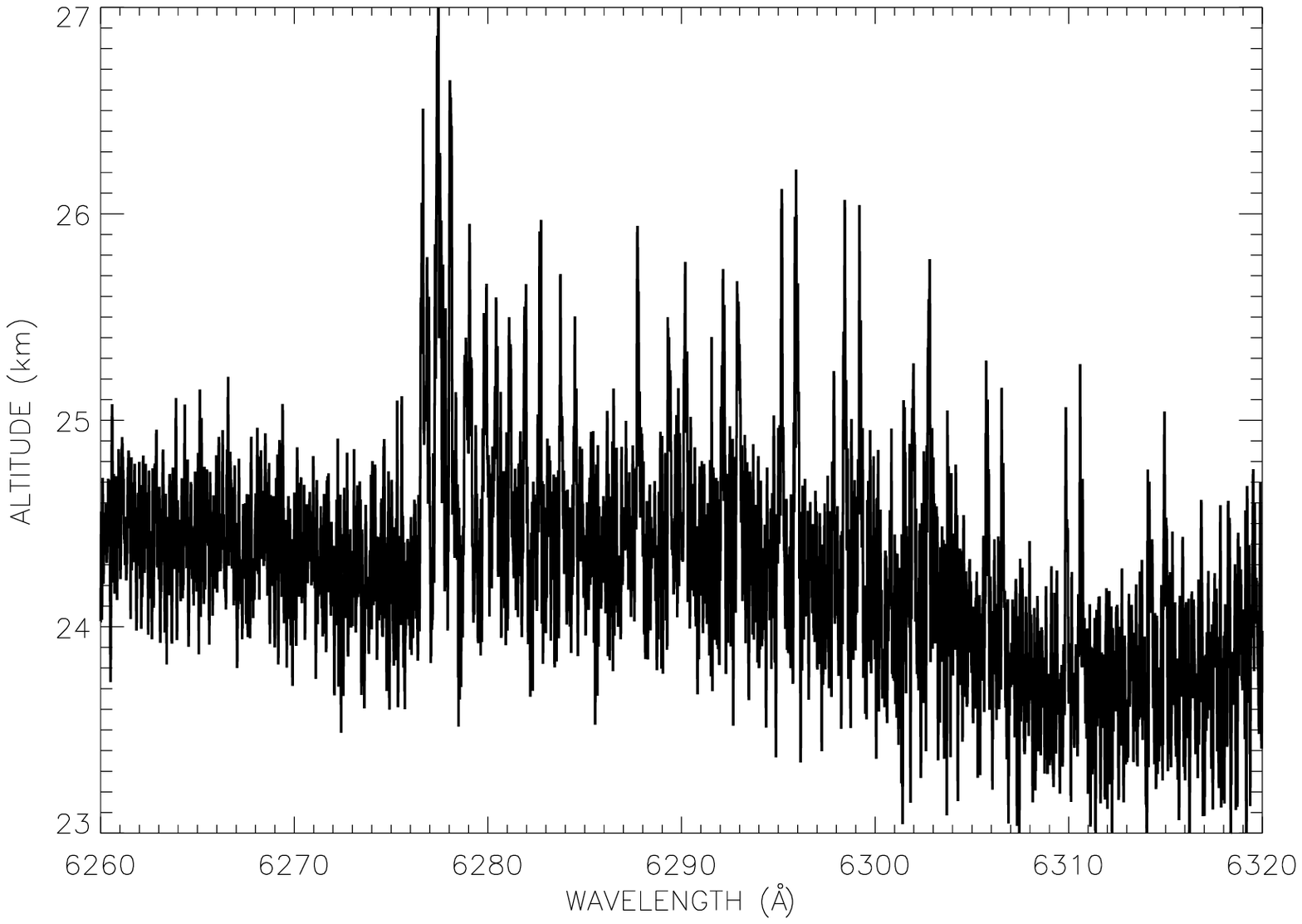}}
\resizebox{\columnwidth}{!}{\includegraphics[width=6cm]{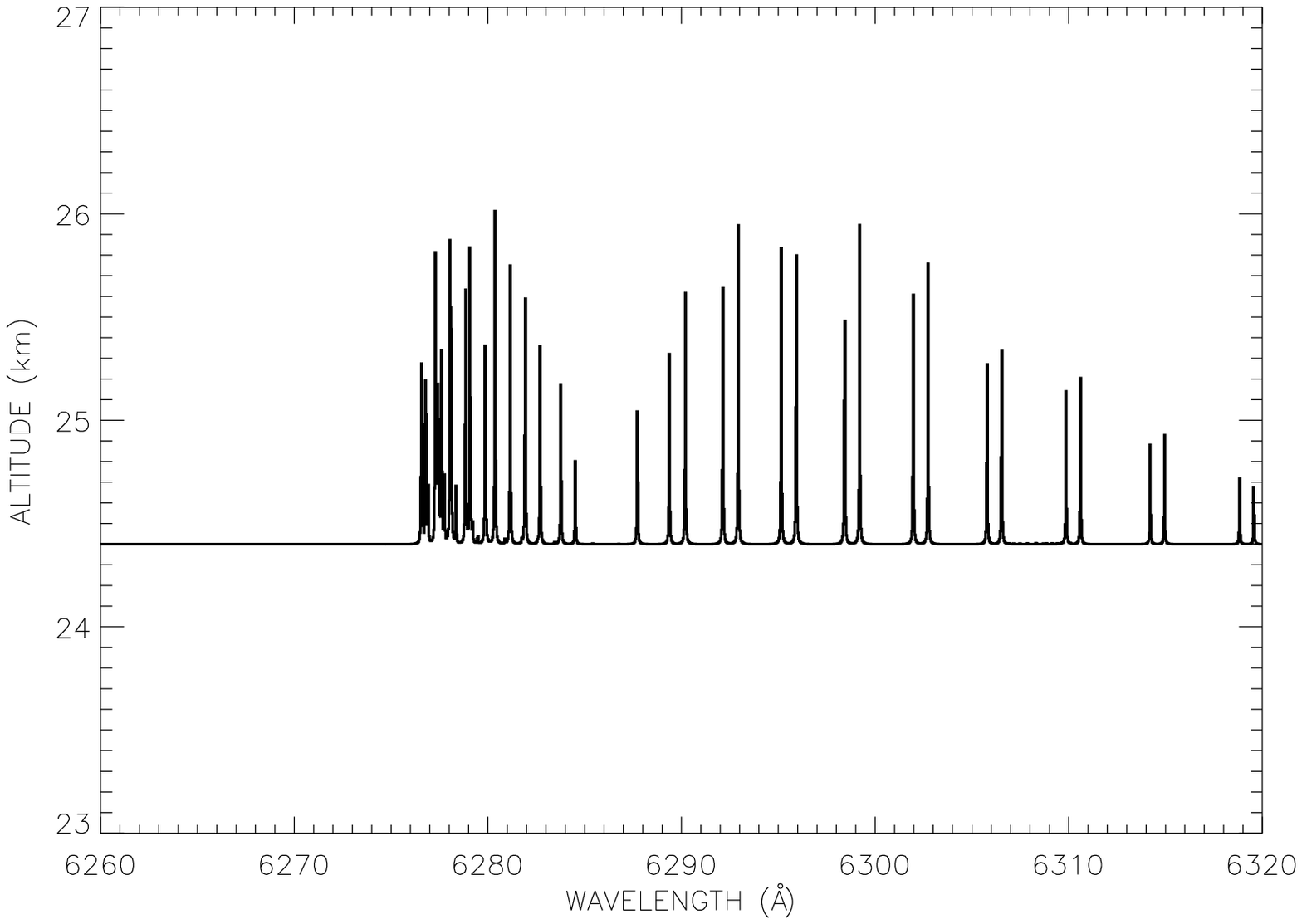}}
   \caption{$\gamma$ (2--0) 6\,280~\AA\ O$_2$ molecular band.
   {\bf Upper plot.} Same background correction
   as in Fig.~\ref{NaI} (see text).
   The solid line represents the altitude information as
   extracted from the SOPHIE data set.
   {\bf Lower plot.}The O$_2$ band spectral structure is shown to
   demonstrate the clear coincidence in position of the repeated
   spikes signing the O$_2$ detection.}
   \label{O2_gamma}
\end{figure}

The other oxygen molecular band in the observed spectral range is
the $\gamma$ (2--0) at 6\,280~\AA . It is shown in
Fig.~\ref{O2_gamma}, for which the same zero level corrections
have again been used.

The narrow spike positions of the band are clearly detected. This
band is weaker however and emerges only by $\sim$1~km above the
continuum. A detailed discussion of this weaker signature is again
beyond the scope of this paper, but this second detection of O$_2$
demonstrates the quality and precision of our approach, which
allows us to detect atmospheric signatures just beyond the photon
noise. Indeed the photon noise induces in that region of high
solar flux and high instrument sensitivity only a $\pm$0.5~km
uncertainty over our altitude evaluation.

Note that for this O$_2$ molecular band we have checked the
perfect match of the peak's spectral positions with the predicted
ones while it seems that again their heights drop faster than the
peak's altitude variations evaluated in the model calculation.
This appears to confirm that the ``standard'' atmospheric model
selected for that comparison is probably hotter than the observed
atmosphere, particularly in the sampled $\sim$30~km altitude
range. If this were true it may also explain the slight mismatch
found between the predicted and the observed O$_3$ Chappuis band
(see Fig.~\ref{h_model}), in which the model is higher by
$\sim$2~km. This band will certainly be scaled down by some amount
in a lower temperature model, simply because the atmospheric scale
height will be reduced.

\subsection{Previous lunar eclipse studies}

Other observations of the 2008 August 16 lunar eclipse were
completed through inter-calibrated optical and near-infrared
ground-based observations at the William Herschel and Nordic
Optical Telescopes by Pall\'e et~al. (2009), to provide continuous
wavelength coverage from 0.36 to 2.40 $\mu$m. Their approach was
very similar to ours, except that their spectral resolution is
somewhat lower ($\sim$1\,000 instead of $\sim$75\,000). They also
focused on umbral, penumbral, and out-of-the-eclipse observations
without detailing (as we did) the precise location within the
penumbra from where the observations were made. This is important
because it leads toward only qualitative detections instead of
quantitative ones.

Furthermore, the Pall\'e et~al. (2009) transmission spectrum is
``calculated by computing the ratio of the umbra/penumbra regions
taken at the same averaged air mass in order to minimize the local
atmosphere's telluric line variations''. This means that the
observed transmission spectrum they evaluated is not the
transit-like spectrum of the Earth. Indeed the umbra spectra
resulting uniquely from rays deflected by the lower and dense
layers of the atmosphere are unobservable during a real transit of
an extrasolar planet since the corresponding stellar light
deflected through the lower layers of the atmosphere moves away
from the observer line of sight. What Pall\'e et~al. (2009) have
reported are identifications of different atmospheric specie
signatures only present in the lower parts of the atmosphere as
seen within the umbra, but not the real transit signatures, which
are only observable within the penumbra, as we did.

\begin{figure}[h]
   \centering
\resizebox{\columnwidth}{!}{\includegraphics[width=6cm]{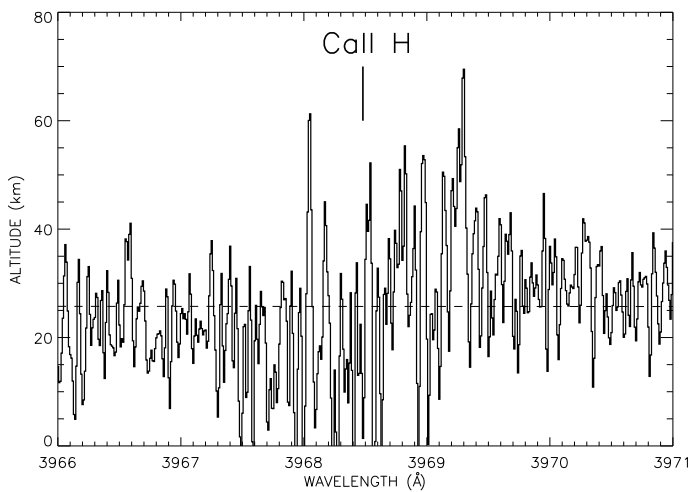}}
   \caption{The $h(\lambda )$ variation as extracted from the region
   of the CaII H line.
   Because the S/N is quite poor in that region the photon noise in this
   part of the spectrum is  more than 10 times higher than in the other
   spectral regions. Large signatures similar to the NaI ones should
   be visible however, but according to the low CaII content of the
   Earth atmosphere, they are simply not observable, as it is the case
   here.}
   \label{CaII}
\end{figure}

In particular, Pall\'e et~al. (2009) have not detected the NaI
signatures. They are produced at very high altitudes only
($\sim$92~km) and should be indeed unobservable in the lunar
umbra. They mention the detection of CaII signatures however,
while it is also known that in the Earth atmosphere CaII ions are
only present at high altitudes (\cite{granier1989}). CaII is
furthermore about 120 times fainter than NaI, thus the CaII
detection (and not the NaI one) is quite a surprise. Because we
have in our hands much lower S/N SOPHIE observations at our
disposal, we tried however to look for possible CaII signatures.
As shown in Fig.~\ref{CaII}, nothing shows up. We conclude that
the Pall\'e et~al. (2009) CaII detection is probably an artifact.

To enforce that argument we note that Pall\'e et~al. (2009)
clearly detected H$_2$O signatures (a low altitude species) at
$\sim$7\,200~\AA , while as shown from model predictions
(\cite{kaltenegger2009}), these could not show up in a transit
spectrum being hidden below the O$_3$ and Rayleigh signatures
(see~Fig.~\ref{h_model}). This is why the different H$_2$O
signatures we have seen in our observation of the ``vertical''
Earth atmosphere (see Fig.~\ref{transmission}) do not show up in
any of our ``horizontal'' transit-like signatures within the Earth
atmosphere, because they are simply too low in altitude as
predicted from the model calculations (see {\it e.g.}
Fig.~\ref{NaI} where several H$_2$O lines should have shown up
around the NaI doublet, if present).

The Pall\'e et~al. (2009) study in a larger spectral domain than
ours reveals the great variety of detectable species, while ours
reveals how precisely the atmospheric content could be analyzed.

\subsection{Detection of O$_2$ and O$_3$ in an extrasolar planet atmosphere during transits}

The purpose of this work is also to show and evaluate the
feasibility of the detection of O$_2$ and O$_3$ in the atmosphere
of an Earth-like extrasolar planet during transits.

We will not repeat the arguments of Lecavelier des Etangs \&\
Ehrenreich (2005) here but only recall the extreme difficulty of
these observations, which demand $10^{-6}-10^{-8}$ accuracy to be
obtained over the observed stellar flux before, during, and after
the planetary transit. This is a real challenge even merely in
terms of photon noise, and asks for very large telescopes in the
10 to 30~m diameter range. This is only possible from the ground
in a relatively short-term perspective, in particular in the frame
of the {\it ELTs} (Extremely Large Telescopes) concepts that are
planned to operate during the coming decade.

The {\it ELTs} are however observing through the Earth atmosphere,
and it is difficult to argue that such high precision accuracies
will be achieved.

Our observations show that although the O$_3$ Chappuis band seems
to be easier to detect according to the broad spectral range over
which it extends, it is still possible to think that uncontrolled
systematics due to atmospheric fluctuations will perturb the
detection. This is the obvious difficulty of broadband detections.

\begin{figure}[h]
   \centering
\resizebox{\columnwidth}{!}{\includegraphics[width=6cm]{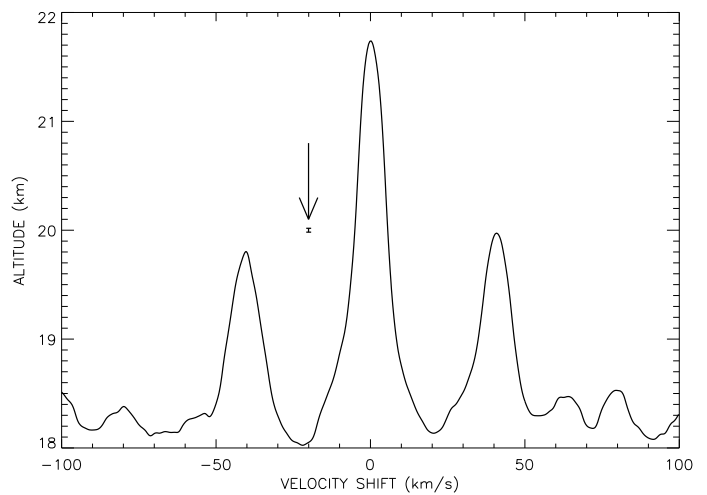}}
  \caption{Correlation function extracted from the
   O$_2$ forbidden $^1\Sigma_g^+$--$^3\Sigma_g^-$
   transition band B (1--0) at 6\,880~\AA\ by using a mask centered on each
   of the spike positions over a $\pm$~0.1~\AA\ width. The typical error bar
   of this function is indicated (arrow) and is on the order of the tracing thickness,
   according to the total spectral width sampled by the mask.}
   \label{O2_peigne}
\end{figure}

On the other hand, our O$_2$ observation via high-resolution
spectroscopy signatures is very promising because it can be shown
that the addition of all spectral lines over only the O$_2$
forbidden $^1\Sigma_g^+$--$^3\Sigma_g^-$ transition band B (1--0)
at 6\,880~\AA\ is equivalent to a detection over a $\sim$8~\AA\
band pass. This places the O$_2$ detection among the feasible ones
from the ground with {\it ELT} telescopes at least in terms of
photon fluxes. But the use of high spectral resolution and stable
spectrographs, similar to the one presently used for radial
velocity searches of extrasolar planets, has proved to be possible
in a quite difficult case, which is to observe the Earth's
atmosphere as a transiting planet through the Earth atmosphere
itself, {\it i.e.} all searched for and perturbing spectral
signatures are exactly at the same wavelengths. For transiting
extrasolar planets the situation will be more favorable because a
simple shift of $\sim$10 to 30~km/s or more will be enough to
clearly separate them as shown through the correlation function
extracted with an O$_2$ like mask (see Fig.~\ref{O2_peigne}). This
should be a relatively common observational situation. Note the
secondary peaks of the correlation function which are due to the
relatively regular separation between the successive O$_2$ line
peaks. Because these are however well separated, this should not
prevent O$_2$ detection over this molecular band.

Our observation shows that with high spectral resolution,
systematics are much easier to control because extremely nearby
spectral regions behave similarly to each other in terms of any
systematic effects. This may be one of the very few possible
approaches to reach this goal, which is detecting O$_2$, one of
the strongest signatures of life bearing atmospheres.

Observing O$_2$ and O$_3$ species in extrasolar Earth like
planetary atmospheres, both accessible from the ground in the same
spectral range, shows that these searches will be attainable for
{\it ELT} observatories equipped with high spectral resolution and
high stability spectrographs.

\section{Conclusion}

We have shown that Lunar eclipse spectral observations are able to
reveal the atmospheric content of the Earth's atmosphere, in
particular, broadband signatures of O$_3$ and Rayleigh scattering
as well as narrowband features of NaI and O$_2$ observed at high
spectral resolution. As these observations mimic future studies of
transiting extrasolar planets, we are confident that quantitative
information about extrasolar atmospheres will be within reach.
Furthermore, both Rayleigh scattering and O$_3$ are broadband and
extend across the visible part of the spectrum, just where
solar-like stars have their maximum flux, making it easier to
obtain the high signal-to-noise values necessary for a positive
exoplanet detection.

More studies of Lunar eclipses should be completed in order to
better quantify the present detections and more precisely show the
feasibility of future extrasolar Earth-like planets studies. These
studies, with extremely large ground-based telescopes, should
allow at least the detections of ozone and O$_2$. In any case,
these observations will be extremely difficult because the
required accuracy is in the $10^{-6}$ to $10^{-8}$ range. This is
why observing in the visible range, where orders of magnitude more
photons are available, could ultimately be one of the most
promising approaches.

\begin{acknowledgements}

The authors thank the staff of Haute-Provence Observatory for
their contribution to the success of the {\it SOPHIE} project and
their support at the 1.93-m telescope. We thank the ``Programme
National de Plan\'etologie'' (PNP) of CNRS/INSU, the Swiss
National Science Foundation, and the French National Research
Agency (ANR-08-JCJC-0102-01 and ANR-NT05-4-44463) for their
continuous support of our planet-search programs.

We also thank W.A. Traub, our referee, for mentioning the ring
effect that possibly explains the zero shift correction and E.
Pall\'e for very constructive discussions before the observing
campaigns of the 2008 August 16 Moon eclipse.

DE acknowledges financial support from the Centre National
d'Etudes Spatiales (CNES) and NCS the support from the European
Research Council/European Community under the FP7 through a
Starting Grant, as well as from the Funda\c{c}\~ao para a
Ci\^encia e a Tecnologia (FCT), Portugal, through a
Ci\^encia\,2007 contract funded by FCT/MCTES (Portugal) and
POPH/FSE (EC), and in the form of grants reference
PTDC/CTE-AST/66643/2006 and PTDC/CTE-AST/098528/2008.

\end{acknowledgements}

\end{document}